\documentclass[
12pt,
reprint,
showpacs,preprintnumbers,
showkeys,
amsmath,amssymb,
aps,
prb,
]{revtex4-1}

\usepackage{graphicx}% Include figure files
\usepackage{dcolumn}% Align table columns on decimal point
\usepackage{bm}% bold math
\usepackage{hyperref}% add hypertext capabilities
\usepackage{subfigure}
\usepackage{nicefrac}
\usepackage{color}
\usepackage{braket}
\usepackage{multirow}
\usepackage{verbatim}

\newcommand{\llangle}{\langle\hspace{-0.4mm}\langle}
\newcommand{\rrangle}{\rangle\hspace{-0.4mm}\rangle}

\newcommand{\lI}{\lambda_{\rm{I}}}
\newcommand{\s}{\sigma}

\newcommand{\La}{\Lambda}
\newcommand{\LI}{\Lambda_{\rm{I}}}

\newcommand{\LBR}{\Lambda_{\rm{R}}}

\newcommand{\LP}{\Lambda_{\rm{PIA}}}

\newcommand{\ii}{\mathrm{i}}

\begin{document}

\title{Copper adatoms on graphene: theory of orbital and spin-orbital effects}

\author{\surname{Tobias} Frank}
\email[Emails to: ]{tobias1.frank@physik.uni-regensburg.de}
\author{\surname{Susanne} Irmer}
\author{\surname{Martin} Gmitra}
\author{\surname{Denis} Kochan}
\author{\surname{Jaroslav} Fabian}

\affiliation{%
 Institute for Theoretical Physics, University of Regensburg,\\
 93040 Regensburg, Germany
 }%

\date{\today}

\begin{abstract}
We present a combined DFT and model Hamiltonian analysis of spin-orbit coupling in graphene induced by copper adatoms in the bridge and top positions, representing isolated atoms in the dilute limit. The orbital physics in both systems is found to be surprisingly similar, given the fundamental difference in the local symmetry. In both systems the Cu $p$ and $d$ contributions at the Fermi level are very similar. Based on the knowledge of orbital effects we identify that the main cause of the locally induced spin-orbit couplings are Cu $p$ and $d$ orbitals. By employing the DFT+$U$ formalism as an analysis tool we find that both the $p$ and $d$ orbital contributions are equally important to spin-orbit coupling, although $p$ contributions to the density of states are much higher. We fit the DFT data with phenomenological tight-binding models developed separately for the top and bridge positions. Our model Hamiltonians describe the low-energy electronic band structure in the whole Brillouin zone and allow us to extract the size of the spin-orbit interaction induced by the local Cu adatom to be in the tens of meV. By application of the phenomenological models to Green's function techniques, we find that copper atoms act as resonant impurities in graphene with large lifetimes of 50 and 100~fs for top and bridge, respectively.
\end{abstract}

\maketitle

%---------------------------------------------------------------
\section{\label{sec:intr} Introduction}
%---------------------------------------------------------------
Adatoms in graphene can fundamentally change the spin properties of graphene,\cite{Neto2009, Han2014} which may bring new advances in spintronics applications.\cite{Zutic2004, Fabian2007} It has been shown experimentally \cite{Han2011, Balakrishnan2013, Gonzalez-Herrero2016} and theoretically,\cite{Yazyev2007, Gmitra2013, Kochan2014} that hydrogen, for example, can induce both local exchange and spin-orbit coupling (SOC), the latter being giant in comparison with the spin-orbit interaction in pristine graphene.\cite{Gmitra2009} These developments point to the possibility of fabricating ultrathin graphene-based magnets or tailored topological materials.

In addition to hydrogen, other adsorbates on graphene have been investigated regarding induced spin properties. It was shown in dedicated density functional theory calculations that the situation with CH$_3$ admolecules closely resembles the one with hydrogen atoms.\cite{Zollner2016} A similar study with fluorine revealed that additionally to the $sp^3$ rehybridization the intrinsic spin-orbit coupling of the fluorine $p$ orbitals is the dominant mechanism enhancing spin-orbit coupling.\cite{Irmer2015}

In terms of magnitudes, the induced spin-orbit coupling can range from 1 meV for hydrogen \cite{Gmitra2013} or CH$_3$, \cite{Zollner2016} through 10 or so meV for fluorine, \cite{Irmer2015, Guzman2015} or even 100 meV for heavy adatoms such as Os,\cite{Hu2012} Au,\cite{Ma2012} Tl, and In \cite{Weeks2011}, which prefer to sit on hollow positions. The heavy adatoms can give rise to topological effects, \cite{Weeks2011, Hu2012} while light adatoms and especially organic molecules (whose presence on graphene is quite likely in ppm concentrations) can lead to resonant scattering and strongly affect resistivity and spin relaxation. \cite{Wehling2010, Hong2011, Kochan2014, Ferreira2014, Tuan2014} So far, there has been no investigation of the induced spin-orbit coupling due to adatoms in the bridge position, presumably as most of the adatoms prefer the top or hollow adsorption sites.\cite{Chan2008}

In recent years Cu adatoms have emerged as important (unintended) functionalization elements, mainly due to the fact that large-scale graphene is grown by chemical vapor deposition (CVD) on Cu substrates. \cite{Mattevi2011} It was shown experimentally (via the spin Hall effect) that CVD grown graphene samples exhibit much lower conductivity and greater spin-Hall angles than exfoliated graphene.\cite{Balakrishnan2014} This all points to a possibly resonant character of the scattering of Dirac electrons in Cu adatoms (or residues), similarly to hydrogen, \cite{Gmitra2013} as well as to a giant induced spin-orbit coupling in graphene due to Cu adatoms. We have earlier predicted for graphene proximitized by the Cu(111) surface, where bonding is only of weak van der Waals type, that it is possible to get large enhancement of spin-orbit coupling. \cite{Frank2016}

The importance of nonlocal interactions to the bonding behavior of coinage metal atoms on graphene was pointed out by Amft et al.,\cite{Amft2011} who studied different approximations to van der Waals interactions and found that within the energy range of meV the bridge and top positions for copper on graphene are energetically equivalent and about 200~meV lower in energy than the hollow position. The energetic equivalency of the adsorption positions reflects in the results of Refs. \onlinecite{Wu2009} and \onlinecite{Cao2010}, which report top and bridge, respectively as lowest energy configuration, without using van der Waals corrections.

Here we explore the spin-orbit coupling effects introduced by single Cu adatoms, taking into account supercells of graphene up to a size of $10 \times 10$ to simulate the dilute limit. We confirm that bridge and top positions are energetically comparable within meV in binding energies. By analyzing the electronic structure, we find that the reduction of symmetry in the bridge case with respect to the top case introduces inequivalence of high symmetry $k$ points, but still yields very similar orbital physics, characterized by $p$ and $d$ contributions at the Fermi energy. We identify the intrinsic spin-orbit interaction of the copper atom as the main source of induced SOC. An analysis of SOC splittings by the usage of the Hubbard $U$ Hamiltonian allows us to quantify the atomic orbital contribution to SOC in terms of Cu $p$ and $d$ orbitals which turn out to be equally important. This proves that Hubbard $U$ corrections can be used as an analysis tool to microscopically understand spin-orbital effects.

Moreover, we derive a new single-orbital tight-binding model Hamiltonian for the bridge system. For the top position we employ the Hamiltonian introduced for hydrogen in Ref. \onlinecite{Gmitra2013}. We show that our model Hamiltonians fit to the low-energy ab-initio data in the whole Brillouin zone. We extract local spin-orbit coupling parameters which are in the order of tens of meV, in agreement with experiment. \cite{Balakrishnan2014} Using the scattering theory with our model Hamiltonians in the dilute limit, we find that copper atoms act as resonant scatterers for both bridge and top positions, again in accordance with experiment. \cite{Balakrishnan2014}
We expect our tight-binding model Hamiltonians to reliably describe the physics near the Fermi level of graphene functionalized with copper and that they can be used in quantum transport simulations that involve orbital and spin-orbital effects, for example spin relaxation, charge and spin transport, or the spin Hall effect. \cite{Ferreira2014, Tuan2014, Bundesmann2015}

This article is structured as follows: We first introduce the employed computational methods and system definition in Sec. \ref{sec:computational_methods}. The electronic structure and the origin of spin-orbit coupling are carefully analyzed in Sec. \ref{sec:results}. After that, we introduce model Hamiltonians for our systems in Sec. \ref{sec:model} where we fit and extract effective spin-orbit coupling strengths. Finally, we apply the model Hamiltonian to analyze the single adatom limit in Sec. \ref{sec:single_adatom}.

%---------------------------------------------------------------
\section{\label{sec:computational_methods} Structure geometry and optimization $\&$ Computational methods}
%---------------------------------------------------------------

To model an isolated copper adatom on graphene, we consider supercells of $5 \times 5$, $7 \times 7$, and $10 \times 10$ units of graphene containing one copper and 50, 98, and 200 carbon atoms, respectively. For orbital effects---binding energies, density of states, atomic, and angular momenta spectral decompositions, and the Bader and L\"{o}wdin charges---we treat the smaller supercells $5\times 5$, and $7\times 7$ which are computationally less demanding.
To study spin-orbit coupling effects  we use a $10\times 10$ supercell to minimize the influence of periodic images.

We used density functional theory (DFT) implemented in the plane wave code \textsc{quantum espresso} \cite{Giannozzi2009} to compute ground state properties of the above specified systems. Our calculations for the graphene supercells were performed at a $k$ point sampling equivalent to a $k$ sampling of $40 \times 40 \times 1$ of a single graphene unit cell. We used ultrasoft Kresse-Joubert\cite{Kresse1999} relativistic PBE\cite{Perdew1996} projector augmented wave (PAW) pseudopotentials. The pseudopotentials incorporate eleven valence electrons for copper and four valence electrons for carbon atoms. A plane wave energy cutoff of 40~Ry and a cutoff of 320~Ry for the Fourier representations of charge density and potential were used. Van der Waals interactions were taken into account by the empirical method Grimme-D2.\cite{Grimme2006} The separating vacuum spacer perpendicular to graphene plane was set to 15\,\AA. Hubbard $U$ corrections\cite{Anisimov1991} were applied for the copper $d$ orbitals in the simplified rotational invariant formulation.\cite{Cococcioni2005}
The initial configuration of a copper atom adsorbed on a specific position on flat graphene was relaxed until the sum of Hellmann-Feynman forces acting on atoms were smaller than $0.001~\rm{Ry}/\rm{a}_0$, using the BFGS algorithm.\cite{Dennis1977} Particularly, the relaxed structure with copper in the top position has a copper-graphene distance of 2.13~{\rm \AA} and the local corrugation---measured as the distance between the graphene plane and the pulled out functionalized carbon atom---of 0.08~{\rm \AA}. For the bridge position we found a copper-graphene distance of 2.16~{\rm \AA} and the local corrugation of 0.11~{\rm \AA}. The small corrugation and relatively large distance to graphene indicates weak bonding. We therefore do not optimize lattice constants and angles as we expect tiny changes of the graphene host system with lattice constant of 2.46~{\rm \AA}, an approach which was already justified in the covalently bonded hydrogen and fluorine cases on the level of $5\times5$ cells. \cite{Gmitra2013, Irmer2015}

%----------------------------
\begin{figure}
	\subfigure[\label{fig:top} top]{
		\includegraphics[width=0.5\columnwidth]{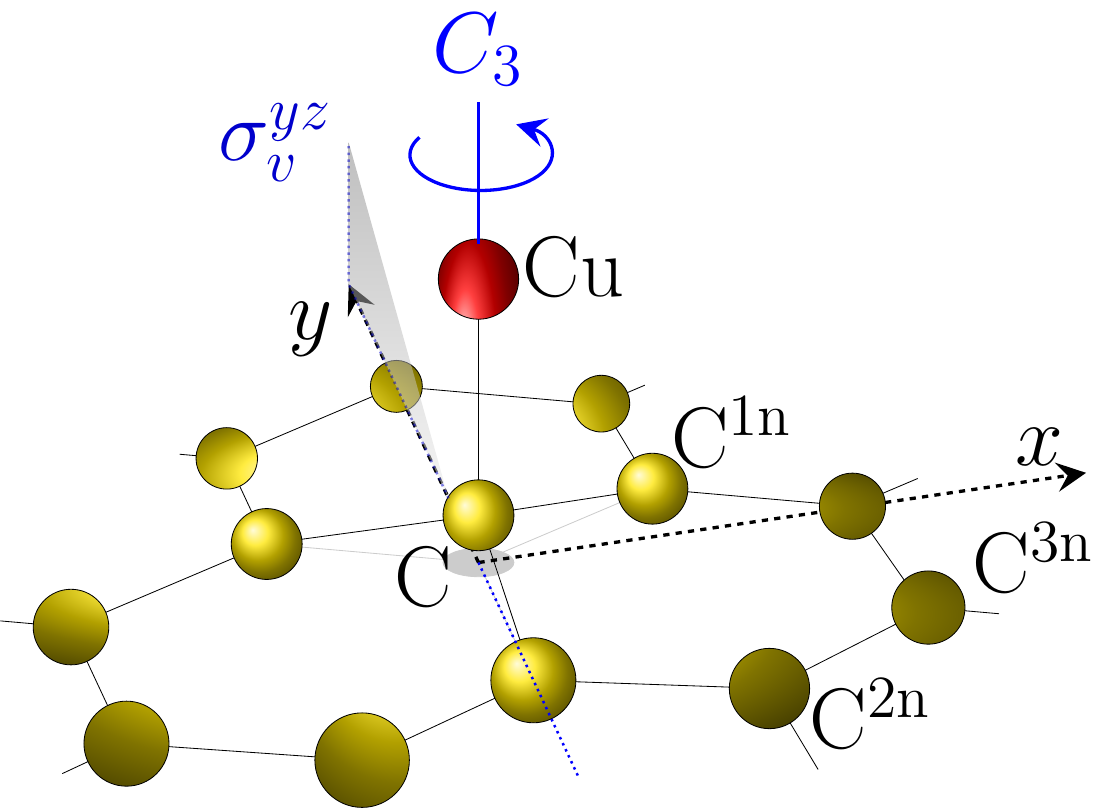}}
	\subfigure[\label{fig:bridge} bridge]{
		\includegraphics[width=0.38\columnwidth]{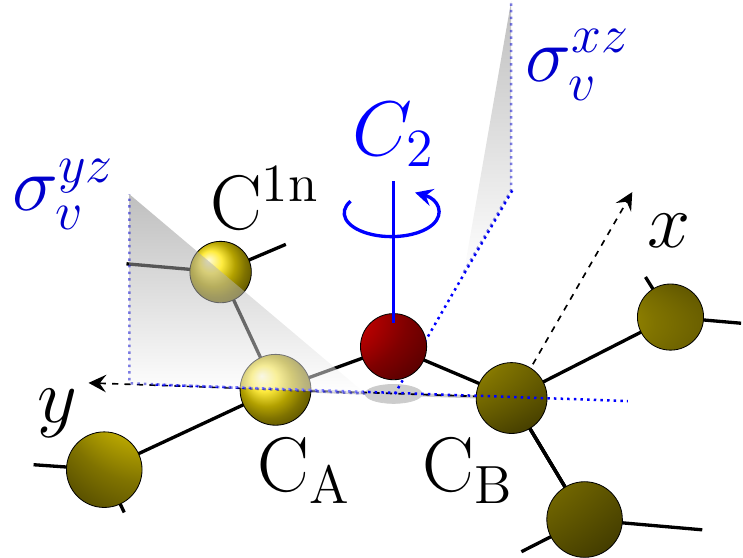}}
	\caption{\label{fig:geometries}
        (Color online) Sketch of the copper adsorption on graphene: (a)~adsorption in the \textbf{top}, (b)~in the \textbf{bridge} position, respectively. Carbon atoms labeling
convention and the local point group symmetry operations are also indicated.}
\end{figure}
%----------------------------

Exchange-correlation functionals including (or supplemented with) van der Waals interactions lead to a significant difference in binding energies of copper on graphene in different adsorption configurations.\cite{Amft2011} Specifically for a $5\times 5$ supercell we found that the top configuration, see Fig. \ref{fig:top}, has about 225~meV lower ground state energy than the hollow configuration, when we use the PBE exchange-correlation functional together with Grimme-D2 van der Waals corrections. Furthermore the top position is just 1~meV below the bridge position, see Fig. \ref{fig:bridge}; therefore from the total energy point of view they can be considered as equal. Hence, in agreement with Ref. \onlinecite{Amft2011} we confirm that the top and bridge configurations---both very close in energy---are more favorable compared to the hollow position. For that reason we focus our analysis on these two configurations.

All our supercells and their reciprocal counterparts possess the full hexagonal geometry. However, they differ by the allowed point group symmetry operations. Namely, the point groups for the top and bridge adatom positions are $C_{3v}$ (6 symmetry operations) and $C_{2v}$ (4 symmetry operations), respectively. This has a direct impact on the shapes of the irreducible wedges that are used to sample the Brillouin zone. For the visualization and local point group symmetries see Fig.~\ref{fig:geometries}.

%---------------------------------------------------------------
\section{\label{sec:results}Electronic Structure --- DFT study}
%---------------------------------------------------------------

We first analyze the orbital electronic structures of Cu in the top and bridge position. The electronic configuration of the outer valence shell of a copper atom is $d^{10} s^1 p^0$. Placing it on graphene, the L\"owdin charge analysis \cite{Lowdin1950} for the copper atom in the top position yields 10.94~e: ($s$, $p$, $d$) = (0.85,  0.26,  9.83)~e, and 11.01~e: ($s$, $p$, $d$) = (0.89,  0.29,  9.83)~e, in the bridge position. One can see that the $s$ and $d$ channels are redistributed and that about 0.3~e resides in the $p$ channel. Alternatively, the Bader charge analysis\cite{Bader1990} unveils that the copper atom has a charge of 10.81~e for the top and 10.75~e for the bridge configuration, respectively. We conclude that the total charge transfer is rather small and copper donates about 0.2 electrons to graphene.

We note that open shell calculations result in a magnetic ground state with magnetic moment of 1~$\mu_\text{B}$ for both studied adsorption configurations, in agreement with results in Refs. \onlinecite{Cao2010, Wu2009}. The total energy gain is about 140~meV compared to the nonmagnetic ground state solution.
The mechanism generating the magnetic state is different from that in hydrogenated graphene where hydrogen also binds in the top position.\cite{Gmitra2013, Kochan2014} In the latter case, the sublattice imbalance of electron occupation of the graphene lattice leads to an extended magnetic moment distribution. Here, it is the unpaired localized $s$ state on the copper that forms the magnetic state,\cite{Cao2010, Wu2009} in a very similar fashion as the copper doublet atomic state $^2S$.

Although there are magnetic ground states, we stick to non-magnetic closed shell calculations as we are interested in spin-orbit coupling effects and want to separate them from potential magnetic effects.

%----------------------------------------------------------------------
\subsection{\label{sec:el_prop_top}Electronic properties---copper in the top position}
%----------------------------------------------------------------------

\begin{figure}
	\begin{minipage}{0.72\columnwidth}
		\includegraphics[width = \columnwidth]{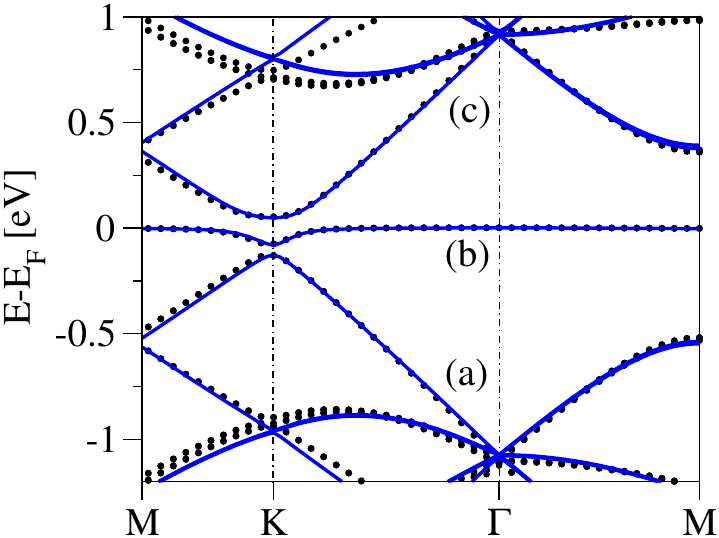}
	\end{minipage}
	\begin{minipage}{0.2\columnwidth}
		\includegraphics[width = \columnwidth]{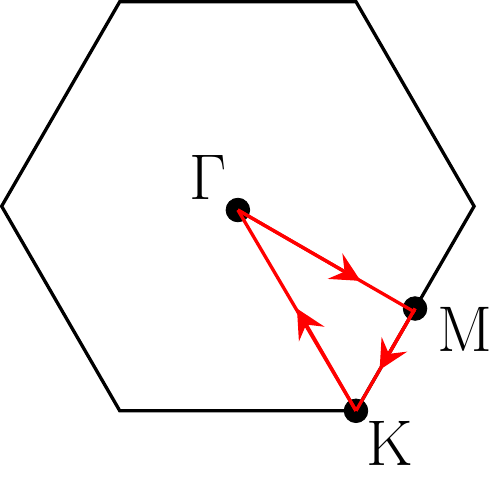}
	\end{minipage}
	\caption{\label{fig:Gr10CuTop_bands}(Color online) Electronic band structure along the high symmetry lines in the first Brillouin zone (sketched at the right) for 10$\times$10
graphene functionalized by copper in the \textbf{top} position. The (black) symbols are first-principles data and the (blue) solid lines correspond to the tight-binding model fit with the hybridization $\omega_t = 0.81\,\mathrm{eV}$ and the on-site energy $\varepsilon_t = 0.08\,\mathrm{eV}$.
Fitting involved the valence (a), midgap (b) and conduction (c) bands around the Fermi level.
		}
\end{figure}

The low energy band structure for a $10\times 10$ supercell of graphene with copper adsorbed in the top position is shown in Fig.~\ref{fig:Gr10CuTop_bands}. Along with the DFT data we present also tight-binding calculated band structure; the model itself is discussed later in Sec.~\ref{sec:top}. Weak bonding of the copper adatom on graphene reflects in the modest binding energy of 0.68~eV and in the residuum of the Dirac cone structure seen in the spectrum. Remnants of the Dirac cone are visible from $-1$ to $0.75$~eV, with respect to the Fermi level.
At the Fermi level there is a flat band, which we call the \textit{midgap} band. The hybridization gap that opens around the K point is the manifestation of the copper-carbon bonding. The Dirac energy, obtained by linearly interpolating bands (a) and (c) to the K point, is situated about 0.1~eV below the Fermi level, i.e. copper acts as a dopant in accordance with the above Bader charge analysis.

%-------------------------------------------------------

\begin{figure}
	\includegraphics[width=0.98\columnwidth]{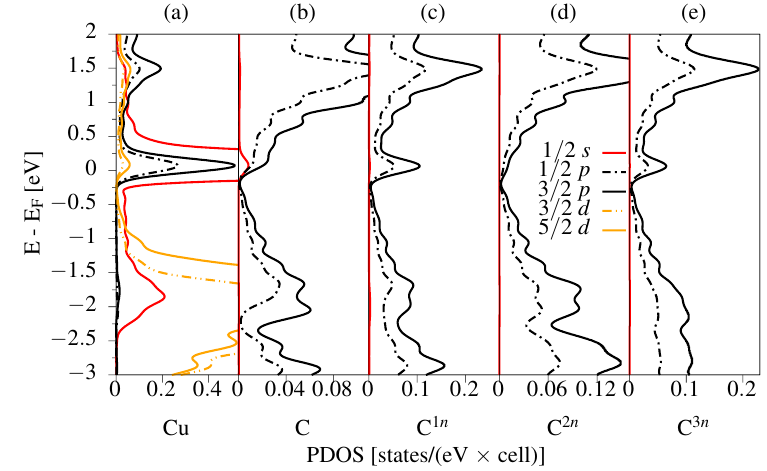}
    \caption{\label{fig:Gr7CuTop_pdos} (Color online) Broadened partial local density of states for $7 \times 7$ graphene supercell with copper adsorbed in the \textbf{top} configuration. (a)~Partial local density of states for the copper adatom, (b)~for the functionalized carbon $\rm C$ and its (c)~nearest-neighbor carbon $\rm C^{1n}$, (d)~second-nearest neighbor carbon $\rm C^{2n}$ and (e)~third-nearest carbon $\rm C^{3n}$. Projected densities are labeled by the total angular momentum $j$ and the corresponding atomic orbital quantum numbers $s$, $p$, $d$, respectively. The numerical broadening is 130 meV.}
\end{figure}

Figure~\ref{fig:Gr7CuTop_pdos} displays the partial local density of states (PLDOS) with the atomic-site-resolved projections on states with different total (orbital + spin) angular momenta. We focus on copper, the functionalized carbon $C$ and its neighboring atoms, see Fig.~\ref{fig:top}. The PLDOS on Cu, see panel (a) in Fig.~\ref{fig:Gr7CuTop_pdos}, is dominated by states with $s$ character near the Fermi level. Small contributions from the $s$ states are also present over the energy range from $-2.5$ to $1.5$~eV. The PLDOS peak at the Fermi level arises from the flat midgap band (b) seen in Fig.~\ref{fig:Gr10CuTop_bands}. The $s$ states of copper play an important role in bonding which can be seen from the hybridization gap in Fig.~\ref{fig:Gr10CuTop_bands} and the overlap in the PLDOS with the electronic states that reside on the neighboring carbon.

The $d$ states extend in the range from $-4$~eV to $-1$~eV with respect to the Fermi level with a maximum contribution at $-2$~eV. The $d$ states of copper with the total angular momentum $j=3/2$ and $j=5/2$ are split in energy by spin-orbit interaction of about $0.2$~eV. This splitting is well understandable in terms of the intra-atomic spin-orbit coupling of the isolated copper whose experimental value is 253~meV.\cite{Sugar1990} However, the weaker intra-atomic spin-orbit splitting of the Cu $4p$ states of 31~meV\cite{Sugar1990} is not visible in the PLDOS.

The PLDOS shows that additionally to copper $s$ states also $p$ and $d$ states are present. For example we find a total $p$ to $d$ ratio of 8.9 for the top case at the Fermi energy. We also analyzed the PLDOS in terms of the orbital angular momentum states, which shows that the DOS around the Fermi energy consists mainly out of $m_z=0$ states (not shown here). However, we find very small contributions of $m_z = \pm 1$, and $\pm 2$ states at the Fermi energy, which should induce spin-orbit coupling.

%-----------------------------------------
%-----------------------------------------

The PLDOSes of the carbon atoms, see Fig.~\ref{fig:Gr7CuTop_pdos}(b)-(e), exhibit approximate linear behavior for electron and hole branches, when ignoring the peaks at the Fermi level. This resembles the linear low energy density of states of pristine graphene and gives a hint for non-invasive and weak bonding of copper to graphene.

%----------------------------------------------------------------------
\subsection{\label{sec:el_prop_bridge}Electronic properties---copper in the bridge position}
%----------------------------------------------------------------------

%-----------------------------------------
\begin{figure}
\begin{minipage}{0.7\columnwidth}
	\includegraphics[width = \columnwidth]{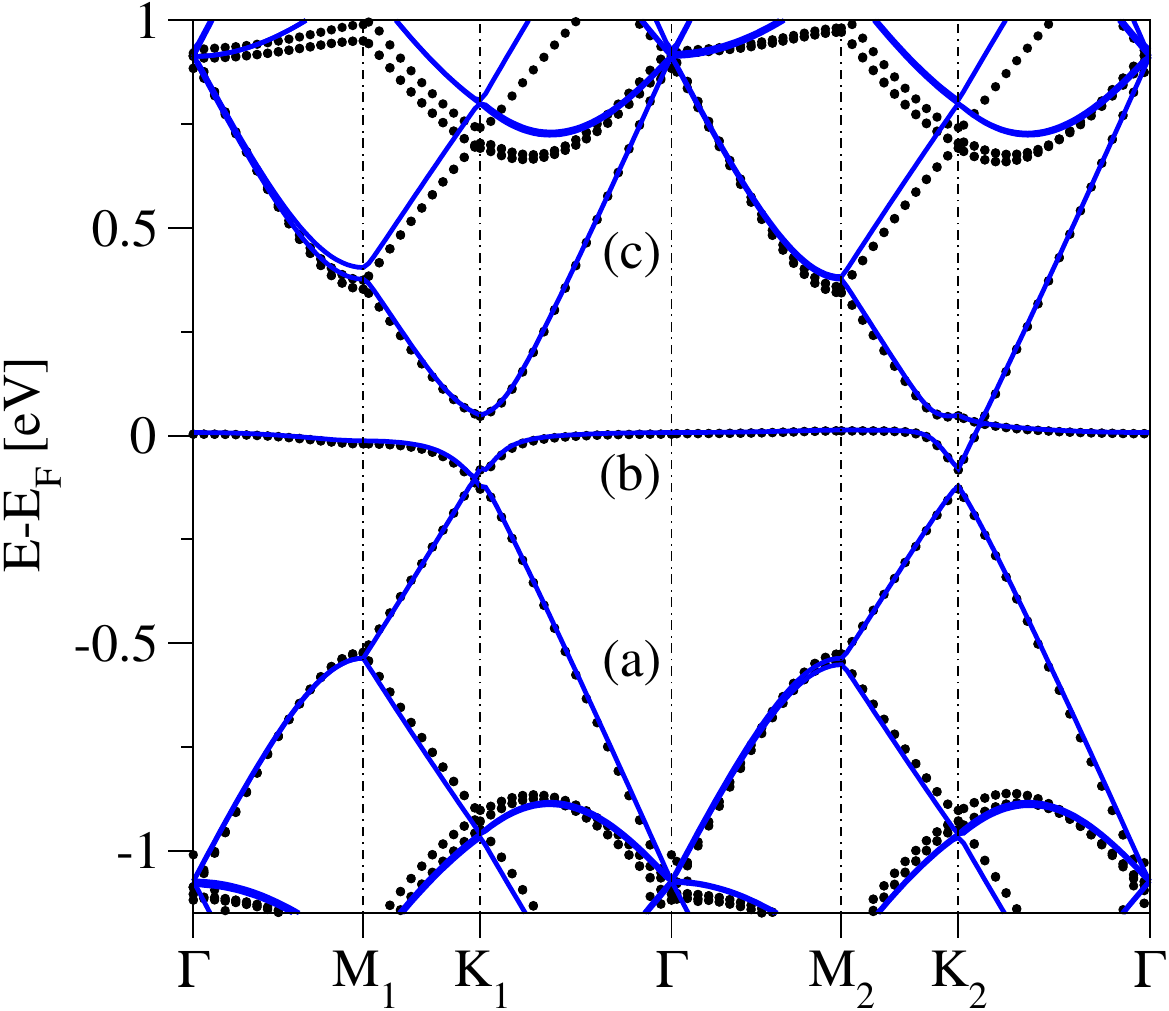}
 \end{minipage}
 \begin{minipage}{0.25\columnwidth}
 	\includegraphics[width = 0.85\columnwidth]{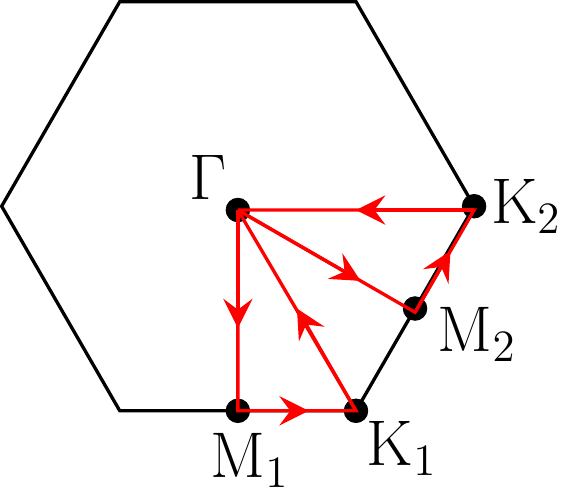}
	\includegraphics[width = 0.6\columnwidth]{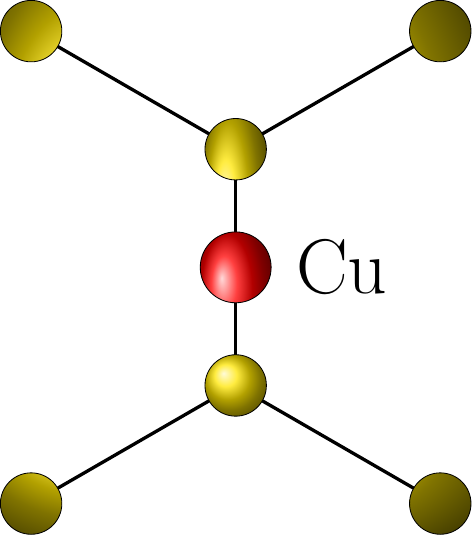}
 \end{minipage}
 \caption{\label{fig:Gr10CuBridge_bands}
(Color online) Electronic band structure along the high symmetry lines in the irreducible wedge of the first Brillouin zone (sketched at the right) for 10$\times$10 graphene functionalized by copper in the \textbf{bridge} position. The (black) symbols are first-principles data and the (blue) solid lines correspond to the tight-binding model fit with the hybridization $\omega_b = 0.54\,\mathrm{eV}$ and the on-site energy $\varepsilon_b = 0.02\,\mathrm{eV}$.
Fitting involved the valence (a), midgap (b) and conduction (c) bands around the Fermi level. Lower sketch at right shows an excerpt of the unit cell around the bridge adatom; coordinate systems of real and reciprocal lattices correspond to each other.
}
\end{figure}
%-----------------------------------------

Figure \ref{fig:Gr10CuBridge_bands} shows the electronic band structure for copper in the bridge position and a sketch of the Brillouin zone including the irreducible wedge---the interior of the trapezoid $\rm{\Gamma M_1 K_1 K_2\Gamma}$. Using time-reversal and translation by a reciprocal lattice vector one can map K$_1$ to K$_2$ and hence the spectrum at those two points should be identical (time-reversal implies only the opposite spin polarization for eigenstates).
This is not the case for M$_1$ and M$_2$ points in the $C_{2v}$ case. There does not exist a transformation combining time-reversal, reciprocal lattice translation and a $C_{2v}$ point group operation that would map M$_1$ to M$_2$, contrary to the $C_{3v}$ case. Therefore the spectra at M$_1$ and M$_2$ are in general distinct. The same holds also for other $k$-points along the high symmetry lines that are displayed in Fig.~\ref{fig:Gr10CuBridge_bands}.
%$\Gamma$K$_1$ and $\Gamma$K$_2$, as well, $\Gamma$M$_1$ and $\Gamma$M$_2$,

To examine those features we have looked at the band structure along the meandering high symmetry path $\rm{\Gamma M_1 K_1 \Gamma M_2 K_2\Gamma}$ inside the
irreducible wedge of the $C_{2v}$ symmetric structure in Fig. \ref{fig:Gr10CuBridge_bands}. We recognize similarities of the band structure compared to the top case. The low energy bands can again be classified in three bands. The difference compared to the top case lies in the observation that along $k$ paths which are perpendicular to the carbon-copper bond (compare $k$ paths $\Gamma \rm{K_2}$ and $\rm{M_1 K_1}$ with the sketch of the local environment of the copper atom in Fig. \ref{fig:Gr10CuBridge_bands}) crossings appear.

%-----------------------------------------
\begin{figure}
 \includegraphics[width=0.98\columnwidth]{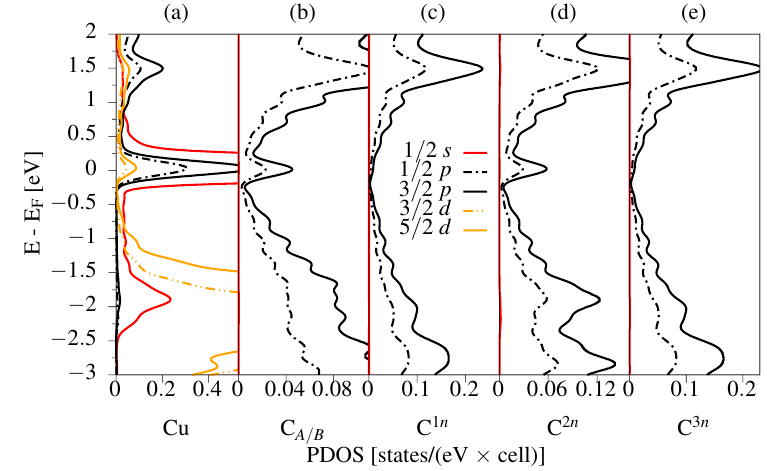}
 \caption{\label{fig:Gr7CuBridge_pdos}
  (Color online) Broadened partial local density of states for $7 \times 7$ graphene supercell with copper adsorbed in the \textbf{bridge} configuration. (a)~Partial local density of states for the copper adatom, (b)~for one (out of two) functionalized carbon $\rm C_{A/B}$ and its (c)~nearest-neighbor carbon $\rm C^{1n}$, (d)~second-nearest neighbor carbon $\rm C^{2n}$ and (e)~third-nearest carbon  $\rm C^{3n}$. Projected densities are labeled by the total angular momentum $j$ and the corresponding atomic orbital quantum numbers $s$, $p$, $d$, respectively. The numerical broadening is 130 meV.}
\end{figure}
%-----------------------------------------

The PLDOS for graphene functionalized by copper at the bridge position, displayed in Fig.~\ref{fig:Gr7CuBridge_pdos} is remarkably similar to the PLDOS analyzed above. Therefore, we qualitatively and quantitatively expect the dominant physical mechanisms for spin-orbit coupling to be the same in both systems. For example, the states with $s$, $p$, and $d$ character appear at the same energies as before, the PLDOS of copper $p$ states at the Fermi energy is 6.9 times larger than the one for copper $d$ states. The total angular momentum states of Cu with $j=3/2$ and $j=5/2$ are again split by $0.2$~eV and the PLDOS peaks near the Fermi level are also built mainly from states with $m_z=0$ (not shown here).
However, there are differences between the two configurations that can be understood in terms of the different underlying point group symmetries. In the bridge case the symmetry group is reduced to $C_{2v}$ and there the concept of high symmetry points and the irreducible wedge in the Brillouin zone differs from the $C_{3v}$ case.

Both copper resolved PLDOSes are very alike which is not surprising given the similar dispersions in Figs.~\ref{fig:Gr10CuTop_bands} and \ref{fig:Gr10CuBridge_bands}. Differences in the binding behavior are most apparent in the density of states of the neighboring carbon atoms. For the top case, the copper $s$ states hybridize with the $\pi$ states of graphene for carbon atoms in the opposite sublattice than the copper atom, which is analogous to hydrogen and fluorine. \cite{Gmitra2013, Irmer2015} In the bridge case, one sees a larger hybridization between copper $s$ states and the $\pi$ states of carbon atom $C_{A/B}$ to which copper binds to.

%-------------------------------------------------------------------
\subsection{\label{sec:origin} Origin of the local spin-orbit coupling}
%-------------------------------------------------------------------

\begin{figure}
 \includegraphics[width=0.98\columnwidth]{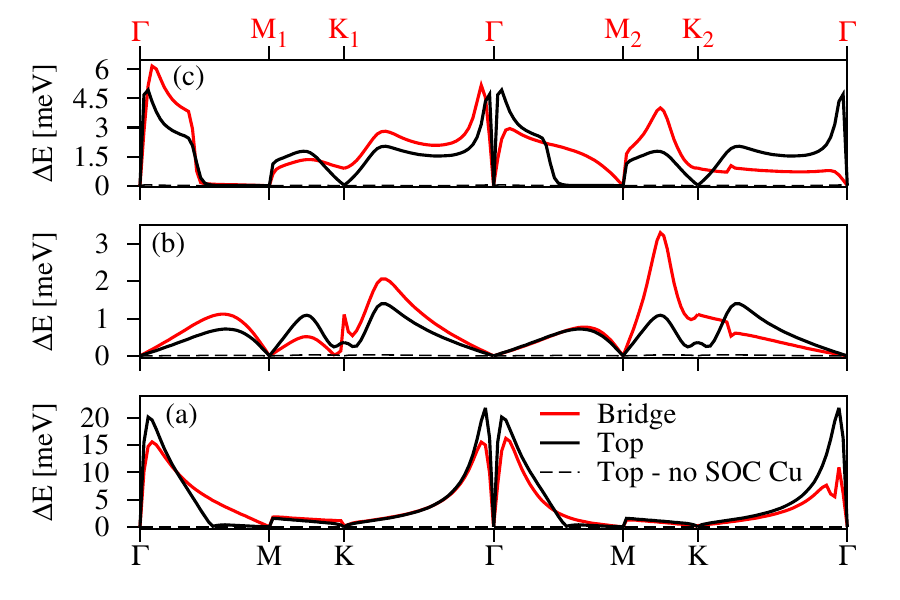}
 \caption{\label{fig:Gr7CuTopBridge_splittings} (Color online) Spin-orbit coupling band splittings for the \textbf{top} (black solid) and \textbf{bridge} copper positions (red solid) for the 7$\times$7 supercell for valence (a), midgap (b) and the conduction (c) band, respectively. Resulting splittings for top case with turned off spin-orbit interaction on copper are shown as well (black dashed). High symmetry points are labeled for the top and bridge cases according to Figs.~\ref{fig:Gr10CuTop_bands} and \ref{fig:Gr10CuBridge_bands}, respectively.
 }
\end{figure}

In order to construct an effective spin-orbit coupling Hamiltonian it is important to analyze its microscopic origin. Figure \ref{fig:Gr7CuTopBridge_splittings} displays the spin-orbit splittings of the valence (a), midgap (b), and conduction (c) bands, respectively, for both the top and bridge adsorption configurations along the indicated high symmetry path for the $7\times 7$ supercell.

The band splittings for the top adsorption show large values up to 20~meV for the valence band, values of 1~meV for the midgap band and up to 4.5~meV in the conduction band. Splittings at $\Gamma$ and M points vanish due to time-reversal symmetry. The midgap band at the K point is still split significantly, though, being mostly lower in the top than in the bridge configuration.

The bridge case is especially interesting because it shows how the spin-orbit coupling splittings are affected by the interactions among the bands. On different segments of the $k$-path in Fig.~\ref{fig:Gr10CuBridge_bands} the spectral repulsion between the midgap and either valence or conduction band has different intensity. Along the path M$_2$K$_2$, where the midgap and conduction bands are closer to each other, the spin-orbit splitting is greatly enhanced to 3~meV as opposed to the path along M$_1$K$_1$ where the spin-orbit coupling is reduced to 0.5~meV.

%------------------------------------------
%------------------------------------------

Figure~\ref{fig:Gr7CuTopBridge_splittings} also shows spin-orbit splittings for the top adsorption configuration when turning off spin-orbit coupling on the copper adatom. Those splittings drop to small values in the range of tens of $\mu$eV and they resemble the spin-orbit splittings calculated for dilute hydrogenated graphene.\cite{Gmitra2013} These residual spin-orbit splittings are due to $sp^3$ hybridization of the carbon atoms in the presence of a local out-of-plane distortion caused by copper, and are negligible. \textit{We conclude that the origin of the local spin-orbit coupling in copper functionalized graphene is due to the intra-atomic spin-orbit coupling of the copper atom.} Note that the DFT band analysis of spin-orbit mediated splittings can give just a qualitative picture as the band splittings are supercell size dependent. The absolute values of spin-orbit coupling strengths can be extracted from a realistic tight-binding model only, as discussed in Sec.~\ref{sec:model}.
	
In what follows we look in more detail on spin-orbit coupling physics between copper and graphene. For practical reasons we take $5\times 5$ supercells to reduce computational costs. We have checked that the orbital decomposition of bands close to the Fermi level changes marginally and hence conclusions drawn from
the smaller supercell analyses are valid also for larger $7\times 7$ and $10\times 10$ supercells.

%------------------------------------------
\begin{figure}
 \includegraphics[width=0.98\columnwidth]{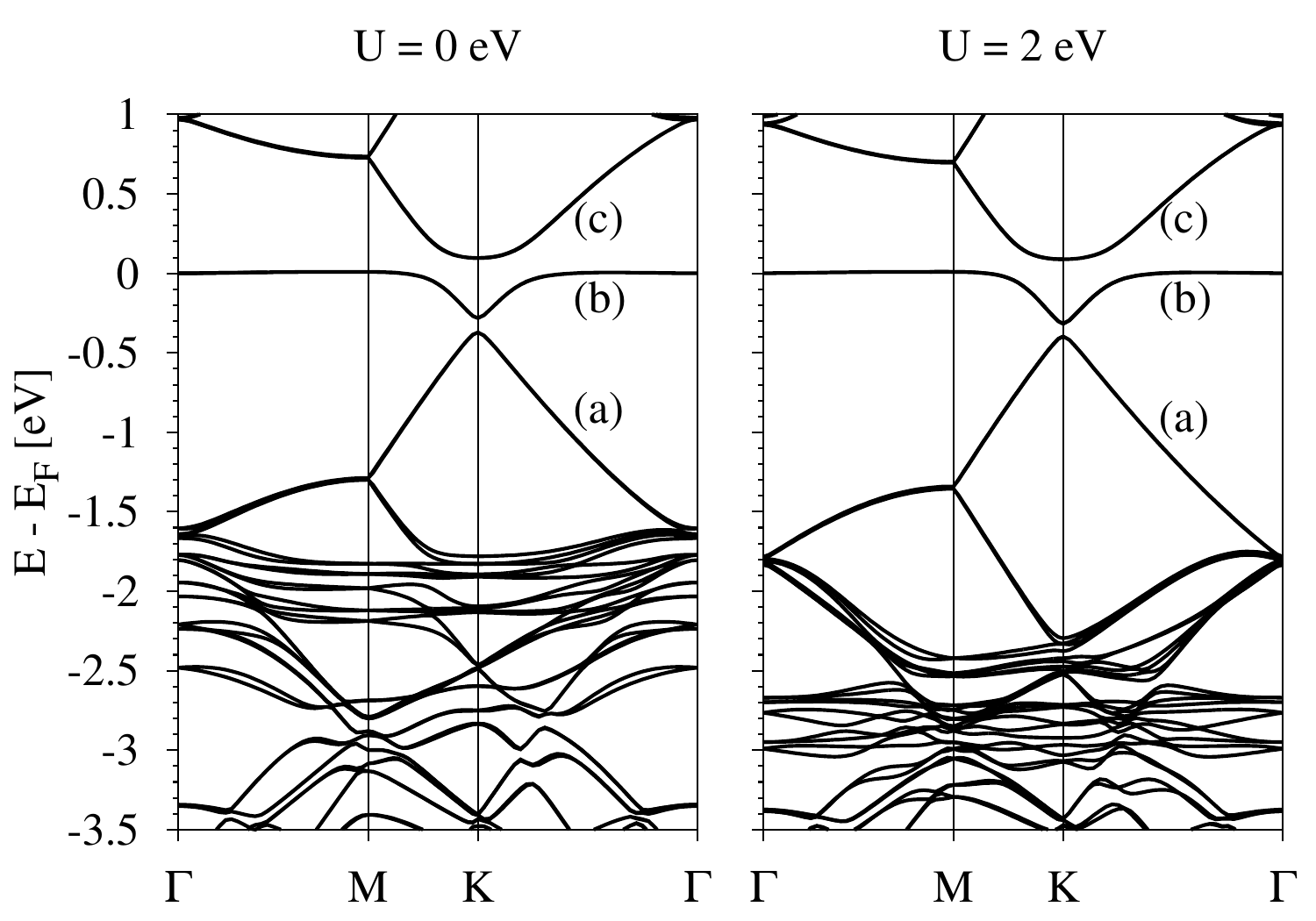}
 \caption{\label{fig:Gr5CuTop_ustudy_bands} Calculated electronic band structure with Hubbard $U$ for $5\times 5$ supercell with copper in the \textbf{top} position. The left panel corresponds to $U=0$\,eV and the right one to $U=2$\,eV. The effect of Hubbard $U$ is clearly seen on the copper $d$ levels which are shifted down in energy for $U=2$\,eV. }
\end{figure}
%------------------------------------------

To separate the spin-orbit effects originating from $d$ and $p$ orbitals, we performed DFT+$U$ calculations.\cite{Anisimov1991} Figure~\ref{fig:Gr5CuTop_ustudy_bands} displays the band structures for copper in the top position for Hubbard $U=0$~eV and $U=2$~eV on the Cu $d$ orbitals. The Hubbard $U$ shifts the fully occupied $d$ states to lower energies. Comparing the left and right panels in Fig.~\ref{fig:Gr5CuTop_ustudy_bands} we see that the shift of the $d$ levels to lower energies starts to modify the band structure from $-1.5$~eV, while near the Fermi level the bands are hardly affected. This is understandable from our previous PLDOS analysis, Fig.~\ref{fig:Gr7CuTop_pdos}: the $d$ level contribution to states near the Fermi level for $U=0$~eV is quite small and their onset at the K point lies at $-1.75$~eV. For Hubbard $U=2$~eV the $d$ state onset is located at $-2.25$~eV.

%------------------------------------------
\begin{figure}
 \includegraphics[width=0.98\columnwidth]{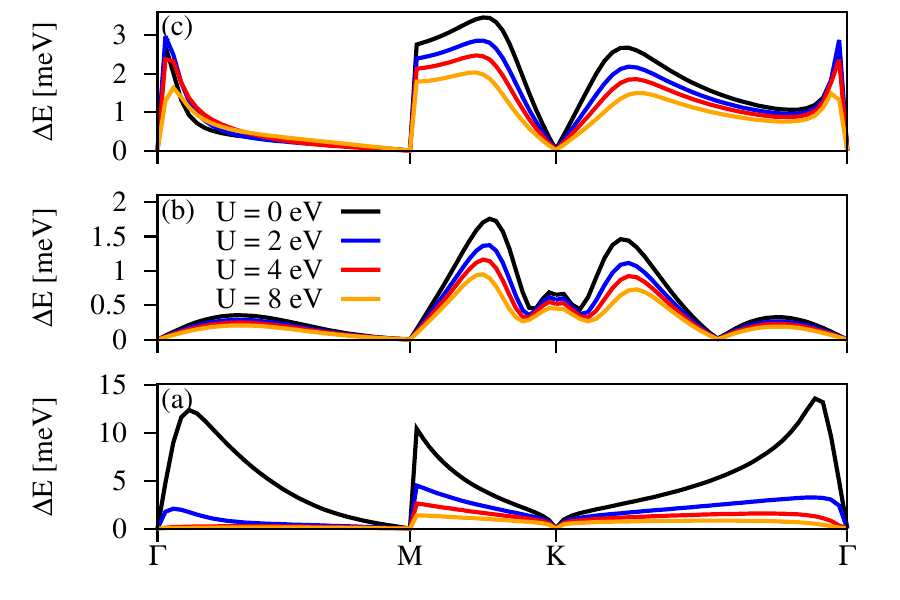}
 \caption{\label{fig:Gr5CuTop_ustudy_splittings} (Color online) Evolution of the spin-orbit band splittings for a $5\times 5$ supercell with copper in the \textbf{top} position for the valence (a), midgap (b) and conduction (c) bands, respectively, with respect to the strength of Hubbard $U$. Different colors correspond to Hubbard $U$ of 0, 2, 4 and 8 eV, respectively.}
\end{figure}
%------------------------------------------

%------------------------------------------
\begin{figure}
 \includegraphics[width=0.88\columnwidth]{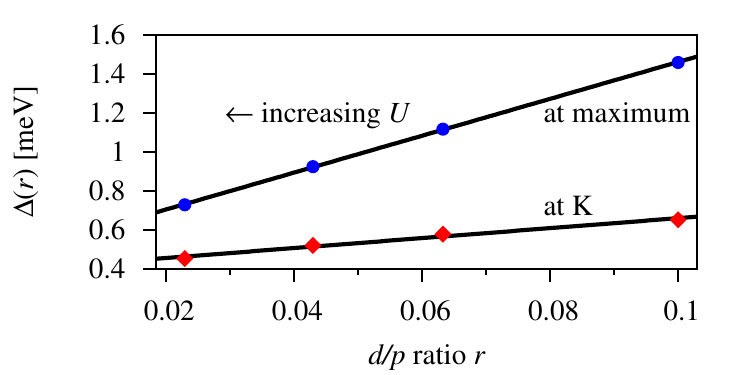}
 \caption{\label{fig:Gr5CuTop_ustudy} Spin-orbit splittings for the midgap band from Fig.~\ref{fig:Gr5CuTop_ustudy_splittings} in the \textbf{top} position, versus the ratio $r=d/p$ between the $d$ and $p$ state densities at the Fermi energy---controlled by the strength of Hubbard $U$. Circles represent the extracted data for the maximal splitting, diamonds represent the extracted data at the K point for $U=0,\,2,\,4,\,8$\,eV and the black lines are linear extrapolations.}
\end{figure}
%------------------------------------------

Figure \ref{fig:Gr5CuTop_ustudy_splittings} shows the spin-orbit coupling band splittings of a copper atom in the top position of the $5\times 5$ supercell for a subsequent series of Hubbard $U=0, 2, 4,$ and 8~eV. Consequently the ratio between $p$ and $d$ density of states at the Fermi level (inside the midgap band) becomes 10.0, 15.8, 23.4, and 43.1, respectively, where the $p$ state contribution remains unchanged. From Fig.~\ref{fig:Gr5CuTop_ustudy_splittings} we see that diminishing the $d$ orbital contributions with raised $U$ decrease the spin-orbit band splittings. This effect is most visible in the valence band since there is higher contribution from $d$ states. The splittings in the midgap and conduction band, however, decrease less drastically. To quantify this behavior we take the maximum of the splitting for the midgap band, $\Delta$, and plot it against the DOS ratio $r=d/p$ at the Fermi level. The graph of $\Delta$ versus $r$ is shown in Fig.~\ref{fig:Gr5CuTop_ustudy}. We see that $\Delta$ scales linearly with $r$ hence writing
\begin{equation}
 \Delta(r) = \tilde{\Delta}\cdot r + \Delta_p\,,
\end{equation}
we can extract $\tilde{\Delta} = 9.5$~meV and $\Delta_p = 0.51$~meV. Extrapolating $\Delta(r)$ for $r\rightarrow 0$, i.e.~for no contributions of $d$ orbitals, one would obtain a splitting of 0.51~meV. Comparing that value with $\Delta(r)$ at $r\simeq0.1$, i.e.~at $U=0$\,eV, we see that ca 35\% of the spin-orbit splitting (at the particular $k$ point and band) is stemming from the $p$ orbitals and 65\% from their $d$ state counterparts. A similar analysis can be carried out for the K point, see Fig. \ref{fig:Gr5CuTop_ustudy}. Here we extract $\Delta_p$ of 0.40~meV. Compared to the $U=0$~eV case with $\Delta=0.65$~meV we find a contribution of 62\% of $p$ orbitals. \textit{Both, $p$ and $d$ orbitals contribute to spin-orbit coupling in nearly equal magnitude.} At first sight it seems quite odd that the maximal splitting at the Fermi level (midgap band) is by 65\% dominated by the $d$ orbitals whose spectral density at this energy is order of magnitude smaller when compared to the $p$ states. But as we already noted, the intra-atomic spin-orbit splitting of $d$ levels of the isolated copper (253\,meV) is order of magnitude larger when compared to $p$ states (31\,meV), so both contributions reasonably compete. This analysis shows that Hubbard $U$ calculations are not just useful for correcting correlations, but can also be used as a tool to better understand microscopic sources of spin-orbit coupling.

%------------------------------------------------------------------------
\section{\label{sec:model} Model}
%------------------------------------------------------------------------

In order to extract realistic parameters for spin-orbit coupling, we construct a model Hamiltonian of the form $\mathcal{H} = \mathcal{H}_0 + \mathcal{H}_\text{orb} + \mathcal{H}_\text{soc}$.
The model Hamiltonian $\mathcal{H}$ accounts for the unperturbed graphene Hamiltonian $\mathcal{H}_0$ and the local perturbation due to copper that has orbital and SOC parts $\mathcal{H}_{\rm orb}$ and $\mathcal{H}_{\rm soc}$, respectively. The unperturbed graphene Hamiltonian has the standard tight-binding form
\begin{align}\label{Eq:Hgraph}
\mathcal{H}_0 = - t &\sum\limits_{\s}\sum\limits_{\left\langle i,j\right\rangle} |c_{i,\sigma}\rangle \langle c_{j,\sigma}| + \nonumber\\
&+
\frac{\text{i}\lI}{3\sqrt{3}}\sum\limits_{\s}{\sum\limits_{\llangle i,j\rrangle}}^\prime
|c_{i,\s}\rangle \nu_{ij} \left(\hat{s}_{z}\right)_{\s\s} \langle c_{j,\s}|\,.
\end{align}
The first term (summation over $\langle.\,,.\rangle$) represents the orbital hoppings among the nearest neighbors\cite{Wallace:PR1947} parametrized by $t=2.6\,\mathrm{eV}$. The second term (summation over $\llangle.\,,.\rrangle$) stands for the SOC mediated spin-conserving hoppings among the second-nearest neighbors\cite{MCCLURE196222} with the intrinsic spin-orbit strength $\lI=12\,\mu$eV.\cite{Gmitra2009} The adatom enhances locally SOC hoppings among specific neighboring carbon sites and those are then excluded from the second summand in Eq.~(\ref{Eq:Hgraph}), indicated by the primed sum symbol. Those omitted SOC contributions appear then in the perturbed SOC Hamiltonian $\mathcal{H}_{\rm soc}$.
Generally, $|c_{i,\s}\rangle$ stands for the carbon $p_z$-orbital with spin $\s$ located at site $i$, the sign symbol $\nu_{ij}$ equals $1(-1)$ depending on whether the second-nearest hopping from $j$ to $i$ via a common neighbor is anticlockwise (clockwise), and $\hat{s}_\alpha$ stands for $\alpha$-th Pauli matrix.

\subsection{\label{sec:top} Top configuration}
\begin{figure}[h!]
\includegraphics[width = 0.7\columnwidth]{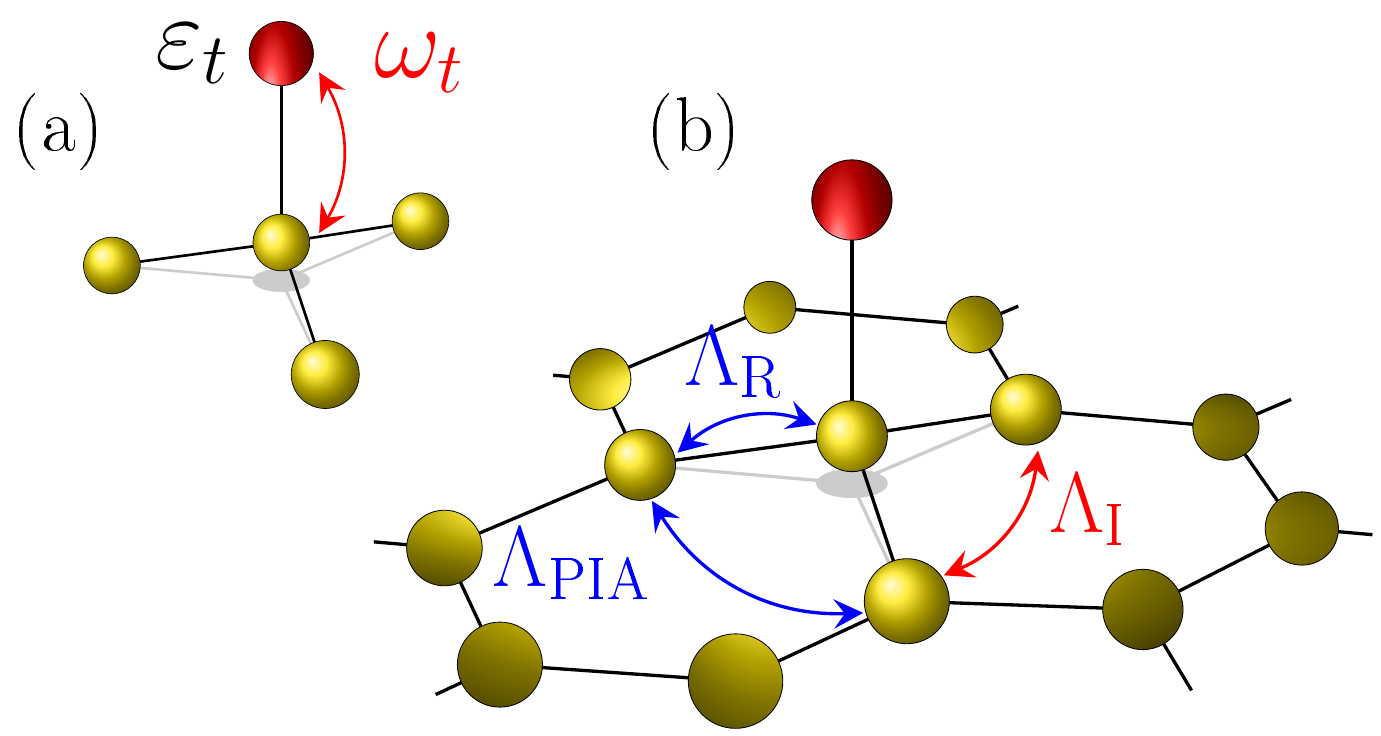}
 \caption{\label{fig:top_orb_soc} (Color online)
 Graphical representation of the minimal orbital and SOC Hamiltonians $\mathcal{H}_{\rm orb}$ and $\mathcal{H}_{\rm soc}$, respectively, for the copper
 in \textbf{top} adsorption position.
 (a)~hybridization hopping $\omega_t$ and the copper on-site energy $\varepsilon_t$,
 (b)~local SOC mediated hoppings among the carbon atoms near the copper $\Lambda_{\rm I}$, $\Lambda_{\rm R}$ and $\Lambda_{\rm PIA}$.
 }
\end{figure}

\paragraph{Orbital Hamiltonian}
The local Hamiltonians $\mathcal{H}_{\rm orb}$ and $\mathcal{H}_{\rm soc}$ describing the monovalent impurity with an effective orbital $|X_\s\rangle$ adsorbed in the top position were already developed in Refs.~\onlinecite{Gmitra2013,Irmer2015,Zollner2016}. Using local atomic orbitals the Hamiltonian $\mathcal{H}_{\rm orb}$ is given as follows:
\begin{align}\label{Eq:HOB_top}
\mathcal{H}_{\rm orb} &= \varepsilon_t \sum\limits_{\s} |X_\s\rangle \langle X_\s|+
\omega_t \sum\limits_{\s} |X_\s\rangle \langle C_\s| + \mathrm{H.c.}\,.
\end{align}
The first term represents the on-site energy $\varepsilon_t$ of an effective copper orbital $|X_\s\rangle$ and the second term parametrized by hopping $\omega_t$ stands for its hybridization with the graphene carbon $|C_\s\rangle$; for the graphical representation see Fig.~\ref{fig:top_orb_soc}(a).
Using the orbital part of $\mathcal{H}_0$ and the local perturbation $\mathcal{H}_{\rm orb}$ one can fit the DFT-computed band structure and extract the values of tight-binding parameters $\varepsilon_t$ and $\omega_t$. Fitting the valence, midgap and conduction bands for the 10$\times$10 supercell shown in Fig.~\ref{fig:Gr10CuTop_bands}, one gets for the top positioned copper $\varepsilon_t=0.08\,\mathrm{eV}$ and $\omega_t=0.81\,\mathrm{eV}$.
The model is quite robust since it allows to excellently fit the three bands along the \textit{complete} MK$\Gamma$M line with only two parameters. Those numerical values are fixed for the following SOC analysis.

%-----------------------------------------------
%-----------------------------------------------

\begin{figure}
 \includegraphics[width = 0.65\columnwidth]{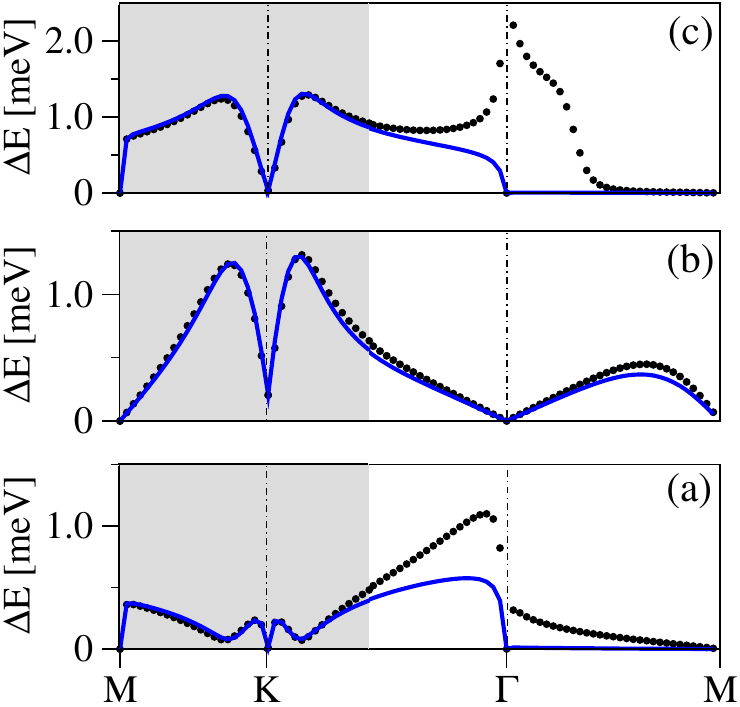}
 \caption{\label{fig:top_soc_split} (Color online) Spin splittings of the valence (a), midgap (b), and conduction band (c), respectively, for the copper on a $10\times 10$ graphene supercell in the \textbf{top} position. First principle data (black symbols) are fitted by the tight-binding model Hamiltonian
 $\mathcal{H}_0+\mathcal{H}_{\rm orb}+\mathcal{H}_{\rm soc}$ for momenta in the shaded regions, the model computed data is represented by solid (blue) lines.}
\end{figure}

\paragraph{Spin-orbital Hamiltonian}
The minimal $C_{3v}$ invariant local SOC Hamiltonian reads,\cite{Gmitra2013,Irmer2015,Zollner2016}
\begin{align}
 &\mathcal{H}_{\rm{soc}} =
 \frac{\text{i}\LI}{3\sqrt{3}} \sum\limits_{\s} \sum\limits_{\llangle i,j\rrangle}
\bigl|C^{\rm{1n}}_{i,\s}\bigr\rangle \nu_{ij} \left(\hat{s}_{z}\right)_{\s\s} \bigl\langle C^{\rm{1n}}_{j,\s}\bigr| \label{Eq:SOC_Hamiltonian}\\
 &+
 \frac{2\text{i}\LBR}{3}\sum\limits_{\s\neq\s'}\sum\limits_{j=1}^3
 \bigl|C_\s\bigr\rangle \left(\bm{\hat{s}}\times\bm{d}_{\mathrm{C},\mathrm{C}^{\rm 1n}_{j}}\right)_{z,\s\s'} \bigl\langle C^{\rm 1n}_{j,\s'}\bigr| + \mathrm{H.c.}\nonumber\\
 &+
 \frac{2\text{i}\LP}{3}\sum\limits_{\s\neq\s'}\sum\limits_{\llangle i,j\rrangle} \bigl|C^{\rm 1n}_{i,\s}\bigr\rangle \left(\bm{\hat{s}}\times\bm{d}_{\mathrm{C}^{\rm 1n}_{i},\mathrm{C}^{\rm 1n}_{j}}\right)_{z,\s\s'}\bigl\langle C^{\rm 1n}_{j,\s'}\bigr|\,.\nonumber
\end{align}
In this case we keep the original SOC terminology introduced in Ref.~\onlinecite{Gmitra2013}: $\LI$ represents the spin-conserving second-nearest neighbor hopping (intrinsic), $\LBR$ the spin-flipping nearest neighbor hopping (Rashba), and $\LP$ the spin-flipping second-nearest neighbor hopping (pseudospin-inversion asymmetry). For a graphical representation of the hoppings see Fig.~\ref{fig:top_orb_soc}(b).
Generally, the symbol $\bm{d}_{\mathrm{C}_{i},\mathrm{C}_{j}}$ stands for the unit vector in the graphene plane that links the annihilation site $\mathrm{C}_j$ with the creation site $\mathrm{C}_i$. Although symmetry allows more local SOC terms in the vicinity of the adatom (see Refs.~\onlinecite{Gmitra2013,Irmer2015,Zollner2016}), we checked that the above three are sufficient to describe the spin-splittings of the bands of interest. Figure~\ref{fig:top_soc_split} shows the fit of the band splittings for the valence, midgap and conduction bands, for the 10$\times$10 system. We restricted our fitting to the low-energy region around the K point (shaded region in Fig.~\ref{fig:top_soc_split}) and obtained $\LI = 9.0\,\mathrm{meV}$, $\LBR = 30.2\,\mathrm{meV}$ and $\LP = -47.4\,\mathrm{meV}$. The orbital and spin-orbital parameters for the top cases of H, F, CH$_3$, and Cu are compiled in Tab. \ref{tab:soc_top}. Comparing copper to fluorine,\cite{Irmer2015} hydrogen,\cite{Gmitra2013} and methyl\cite{Zollner2016} we see that the present SOC parameters are order or two orders of magnitude larger.

%-----------------------------------------------
\begin{table}[htb]
    \caption{\label{tab:soc_top}
        Orbital and SOC tight-binding parameters for adatoms in the \textbf{top} position. In this work, $\Lambda_{\textrm{I}}$ is equivalent to $\Lambda_{\textrm{I}}^{\textrm{B}}$ and $\Lambda_{\textrm{PIA}}$ to $\Lambda_{\textrm{PIA}}^{\textrm{B}}$ of Refs. \onlinecite{Irmer2015, Gmitra2013, Zollner2016}. $\Lambda_{\textrm{I}}^{\textrm{A}}$ is the spin-conserving hopping from the decorated carbon orbital to its next-nearest neighbors.
    }
    \begin{ruledtabular}
        \begin{tabular}{lccccccr}
            Atom&
            \textrm{$\omega_t$[eV]}&
            \textrm{$\varepsilon_t$[eV]}&
            \textrm{$\Lambda_{\textrm{I}}^{\textrm{A}}$[meV]}&
            \textrm{$\Lambda_{\textrm{I}}^{\textrm{B}}$[meV]}&
            \textrm{$\Lambda_{\textrm{PIA}}^{\textrm{B}}$[meV]}&
            \textrm{$\Lambda_{\textrm{R}}^{\textrm{}}$[meV]}\\
            \colrule
            H\cite{Gmitra2013}& 7.5 & 0.16 & -0.21 & - &-0.77 & 0.33\\
            F\cite{Irmer2015}& 5.5 & -2.2 & - & 3.3 & 7.3 & 11.2\\ 
            CH$_3\cite{Zollner2016}$& 7.6 & -0.19 & -0.77 & 0.15 & -0.69 & 1.02\\
            Cu& 0.81 & 0.08 & - & 9.0 & -47.4 & 30.2\\
        \end{tabular}
    \end{ruledtabular}
\end{table}
%-----------------------------------------------

Along the full MK line and about one-third of K$\Gamma$ line the model excellently reproduces the DFT data. Approaching the $\Gamma$ point the model strongly deviates for the valence and conduction bands from first principles, but still stays perfectly aligned for the midgap band. This is because at the $\Gamma$ point the valence and conduction bands lie far away from the Fermi level and other states contribute with different angular momenta (see Fig. \ref{fig:Gr7CuTop_pdos}). Our effective low energy Hamiltonian assumes that all participating atomic orbitals transform with respect to $C_{3v}$ as states with $m_z=0$, which ceases to hold far away from the Fermi level.

%-----------------------------------------------
%-----------------------------------------------

%-----------------------------------------------
\subsection{\label{sec:bridge} Bridge configuration}
%-----------------------------------------------

%-----------------------------------------------
\begin{figure}
 \includegraphics[width = 0.75\columnwidth]{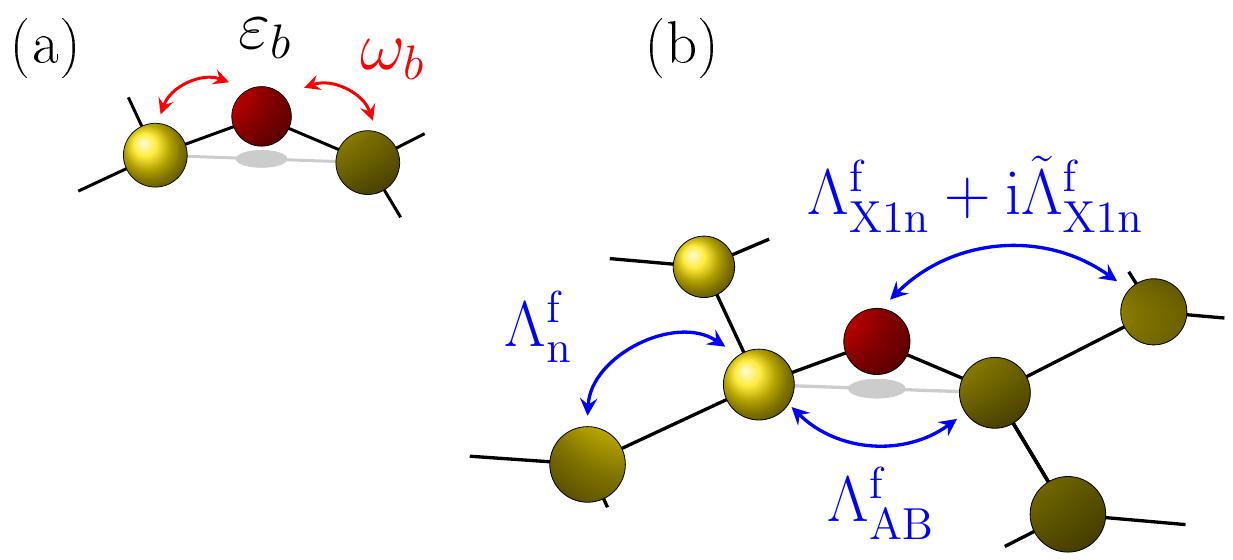}
\caption{\label{fig:bridge_orb_soc}(Color online)
 Graphical representation of the minimal orbital and SOC Hamiltonians $\mathcal{H}_{\rm orb}$ and $\mathcal{H}_{\rm soc}$, respectively, for the copper
 in \textbf{bridge} adsorption position.
 (a)~hybridization hopping $\omega_b$ and the copper on-site energy $\varepsilon_b$,
 (b)~local SOC mediated hoppings among the carbon atoms near copper $\Lambda_{\rm n}^{\rm f}$, $\Lambda_{\rm AB}^{\rm f}$, and $\Lambda_{\rm X1n}^{\rm f}+\ii
 \tilde{\Lambda}_{\rm X1n}^{\rm f}$.
 }
\end{figure}
%-----------------------------------------------

The local point group symmetry for the adatom binding in the bridge position is $C_{2v}$. The structure remains invariant under $C_2$ rotation around the principal axis and the vertical reflections $\sigma_v^{\rm xy}$ and $\sigma_v^{\rm xz}$; see Fig.~\ref{fig:bridge} for the symmetry operations and site labeling. In what follows, we focus on the energy region around the Fermi level which mainly comprises atomic states with $m_z=0$ character. Therefore, to construct the effective low-energy Hamiltonians $\mathcal{H}_{\rm{orb}}$ and $\mathcal{H}_{\rm{soc}}$ we take local atomic orbitals with $m_z=0$ angular momentum as a basis. Knowing their transformational properties under time-reversal and $C_{2v}$ symmetry we can combine them in a $C_{2v}$-invariant way to get $\mathcal{H}_{\rm{orb}}$ and $\mathcal{H}_{\rm{soc}}$. We consider an effective copper orbital $|X\rangle$, the $p_z$-orbitals on two functionalized carbon atoms $|C_{\rm A}\rangle$ and $|C_{\rm B}\rangle$, and the $p_z$-orbitals on their first nearest neighbors $|C^{\mathrm{1n}}_j\rangle$ to describe the local influence of copper.

\paragraph{Orbital Hamiltonian}
A direct generalization of Eq.~(\ref{Eq:HOB_top}) becomes,
\begin{align}\label{Eq:HOB_bridge}
\mathcal{H}_{\rm orb} = \varepsilon_{b} &
\sum\limits_{\s} |X_{\s}\rangle \langle X_{\s}|\nonumber\\
&+\omega_b \left(\sum\limits_{\s}|X_\s\rangle \langle C_{\mathrm{A}\s}| +
|X_\s\rangle \langle C_{\mathrm{B}\s}|\right)+ \mathrm{H.c.}\,.
\end{align}
Again, the first term stands for the on-site energy offset of the copper $|X\rangle$ orbital with respect to $p_z$-carbon levels, and the second one gives the
hybridization between $|X\rangle$ and the functionalized carbon orbitals $|C_{\rm A}\rangle$ and $|C_{\rm B}\rangle$, respectively. For the graphical representation of $\mathcal{H}_{\rm orb}$ see Fig.~\ref{fig:bridge_orb_soc}(a). Ignoring the SOC part of $\mathcal{H}_0$, Eq.~(\ref{Eq:Hgraph}), we can fit $\mathcal{H}_0+\mathcal{H}_{\rm orb}$ with respect to the DFT computed band structure. As a result we obtain
$\varepsilon_b = 0.02\,\mathrm{eV}$ and $\omega_b = 0.58\,\mathrm{eV}$. The comparison between the ab-initio and tight-binding calculations is shown in Fig.~\ref{fig:Gr10CuBridge_bands}. The orbital tight-binding model is quite robust; it allows to perfectly fit the three bands around the Fermi level along the complete $\rm{\Gamma M_1 K_1 \Gamma M_2 K_2\Gamma}$ path inside the irreducible wedge with only two parameters.

\paragraph{Spin-orbital Hamiltonian}
The SOC Hamiltonian in the $C_{2v}$ case is much richer than in the $C_{3v}$ case since the reduced symmetry allows more local hoppings which are themselves represented by complex-valued
SOC strengths. We checked several combinations and in what follows we present a minimal SOC Hamiltonian $\mathcal{H}_{\rm soc}$ able to reproduce the observed spin-orbit splittings around the Fermi level. In the Hamiltonian $\mathcal{H}_{\rm soc}$ given below, all $\Lambda$s are real-valued, and its form relies
on the geometry and chosen axis orientations as shown in Fig.~\ref{fig:bridge}:
\begin{widetext}
\begin{align}\label{Eq:HSOB}
\mathcal{H}_{\rm soc} &=
\mathrm{i}\La^{\mathrm{f}}_{\rm{AB}}\sum\limits_{\s\neq\s'}\sum\limits_{ \langle i, j\rangle}
\bigl|C_{\mathrm{A},\s}\bigr\rangle \bigl\langle C_{\mathrm{B},\s'}\bigr|
\,\bigl[\hat{s}_x\bigr]_{\s\s'} + \mathrm{H.c.}\nonumber\\
&+\La^{\rm{f}}_{\rm{n}} \sum\limits_{\s\neq\s'}\left(
\sum\limits_{\langle\mathrm{A}, j\rangle}
\bigl|C_{\mathrm{A},\s}\bigr\rangle \bigl\langle C_{j,\s'}^{\mathrm{1n}}\bigr|\,
-
\sum\limits_{\langle\mathrm{B}, j\rangle}
\bigl|C_{\mathrm{B},\s}\bigr\rangle \bigl\langle C_{j,\s'}^{\mathrm{1n}}\bigr|\right)
\nu_{\mathrm{X},\mathrm{C}_{j}^{\mathrm{1n}}}\,\bigl[\hat{s}_y\bigr]_{\s\s'}\,
+\mathrm{H.c.}\\
&+
\sum\limits_{\s\neq\s'}\left(
\sum\limits_{\langle\mathrm{B}, j\rangle}
\bigl|X_\s\bigr\rangle \bigl\langle C_{j,\s'}^{\mathrm{1n}}\bigr|\,
-
\sum\limits_{\langle\mathrm{A}, j\rangle}
\bigl|X_\s\bigr\rangle \bigl\langle C_{j,\s'}^{\mathrm{1n}}\bigr|\right)
\,\Bigl\{\nu_{\mathrm{X},\mathrm{C}_{j}^{\mathrm{1n}}}\,\bigl[\ii\,\hat{s}_y\bigr]_{\s\s'}\,\La_{\rm X1n}^{\rm f}+\mathrm{i}\,\tilde{\La}_{\rm X1n}^{\rm f}\Bigr\}+\mathrm{H.c.}\,.\nonumber
\end{align}
\end{widetext}
The first term, parametrized by $\La^{\mathrm{f}}_{\rm{AB}}$, represents the spin-flipping hopping between the functionalized carbon sites $\rm C_A$ and $\rm C_B$. The second term, represented by summation over $\langle \mathrm{A(B)},j\rangle$, accounts for the spin-flipping hoppings between the given functionalized carbon $\rm C_A$($\rm C_B$) and its two nearest neighbors $\rm C^{1n}$, see Fig.~\ref{fig:bridge}. Those hoppings are parametrized by SOC strength $\La^{\rm{f}}_{\rm{n}}$. Symbol $\nu_{\mathrm{X},\mathrm{C}^{\mathrm{1n}}}$ has the same meaning as in Eq.~(\ref{Eq:Hgraph}), assuming the common neighbor of X=Cu and $\rm C^{1n}$ is the functionalized carbon between them. The third line represents the spin-flipping hoppings between the copper $|X\rangle $ and the four $|C^{\rm 1n}\rangle$ orbitals, again see Fig.~\ref{fig:bridge}. It is parametrized by the complex-valued parameter $\La_{\rm X1n}^{\rm f}+\mathrm{i}\,\tilde{\La}_{\rm X1n}^{\rm f}$. The second-nearest neighbor summation over $\llangle \mathrm{X}, j \rrangle$, which would naturally emerge there, was split into the two nearest-neighbor summations $\langle \mathrm{B(A)}, j\rangle$. Graphical representation of the above defined SOC strengths is displayed in Fig.~\ref{fig:bridge_orb_soc}(b).

\begin{figure}
 \includegraphics[width = 1.1\columnwidth]{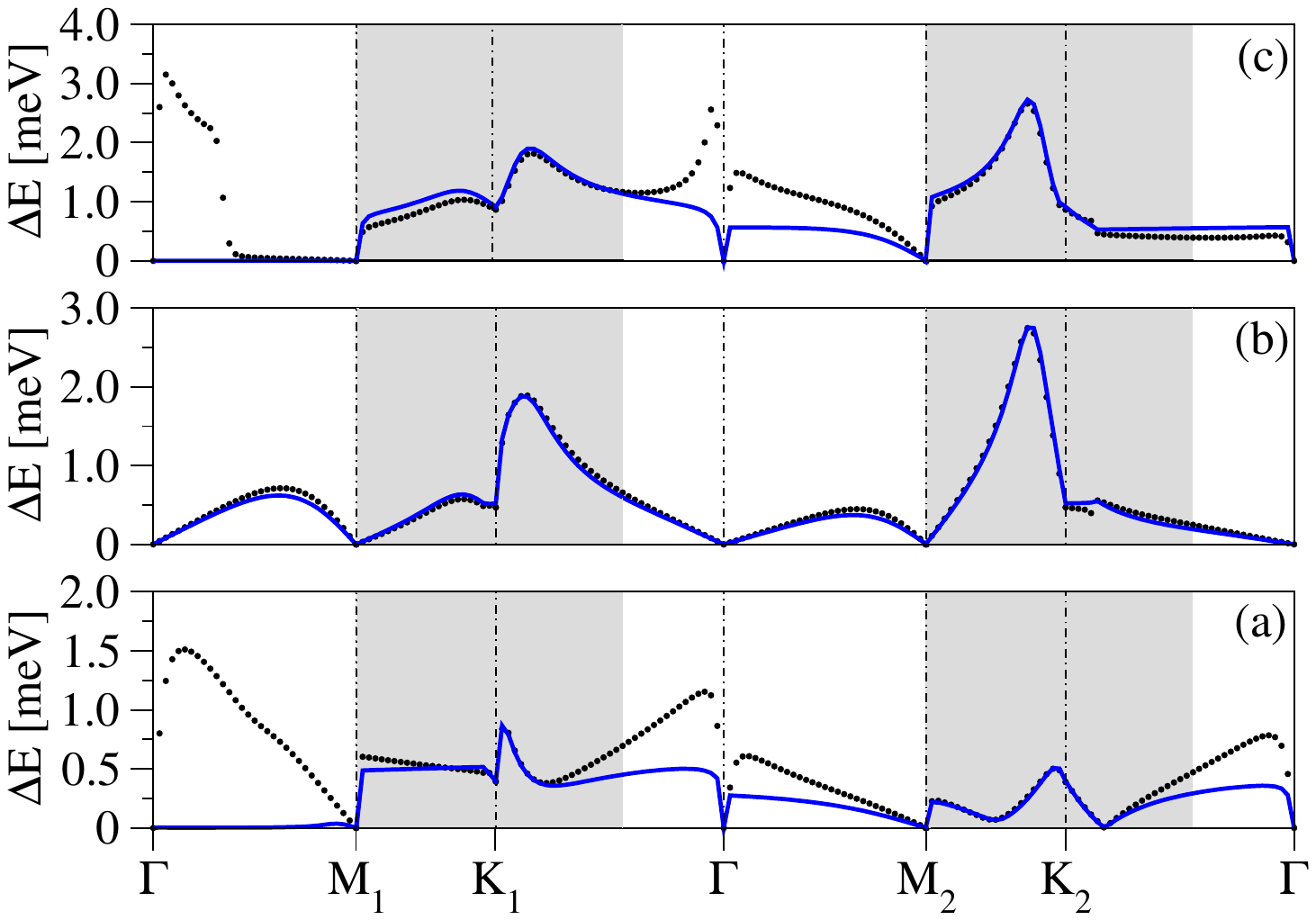}
 \caption{\label{fig:bridge_soc_split} (Color online) Spin splittings of the valence (a), midgap (b), and conduction band (c), respectively, for the copper atom on $10\times 10$ graphene supercell in the \textbf{bridge} position. First principle data (black symbols) are fitted by the tight-binding model Hamiltonian
 $\mathcal{H}_0+\mathcal{H}_{\rm orb}+\mathcal{H}_{\rm soc}$ for momenta in the shaded regions, the model computed data is represented by solid (blue) lines.}
\end{figure}

Figure~\ref{fig:bridge_soc_split} shows the fit of the band splittings for the valence, midgap and conduction bands. We use the full bridge model Hamiltonian $\mathcal{H}_0+\mathcal{H}_{\rm orb}+\mathcal{H}_{\rm soc}$ and fit the low-energy regions around the K$_1$ and K$_2$ points, respectively, (shaded region in Fig.~~\ref{fig:bridge_soc_split}). We obtain the following values for the local SOC parameters:
$\La^{\mathrm{f}}_{\rm{AB}} = 41\,\mathrm{meV}$, $\La^{\rm{f}}_{\rm{n}}= -7.5\,\mathrm{meV}$,
$\La^{\rm f}_{\rm{X1n}}= 1.4\,\mathrm{meV}$, $\tilde{\La}^{\rm f}_{\rm{X1n}}= 8.4\,\mathrm{meV}$.
Approaching the $\Gamma$ point the model again deviates from first principles for the valence and conduction band.
The reason is obvious: Close to the $\Gamma$ point and at energies away from the Fermi level there dominantly contribute states with $m_z\neq 0$. These are not included in our effective low energy model. The parameters are summarized in Tab. \ref{tab:soc_bridge}.

%-----------------------------------------------
\begin{table}[htb]
    \caption{\label{tab:soc_bridge}
        Orbital and SOC tight-binding parameters for Cu in the \textbf{bridge} position.
    }
    \begin{ruledtabular}
        \begin{tabular}{ccccccr}
            \textrm{$\omega_b$[eV]}&
            \textrm{$\varepsilon_b$[eV]}&
            \textrm{$\Lambda_{\textrm{AB}}^{\textrm{f}}$[meV]}&
            \textrm{$\Lambda_{\textrm{n}}^{\textrm{f}}$[meV]}&
            \textrm{$\Lambda_{\textrm{X1n}}^{\textrm{f}}$[meV]}&
            \textrm{$\tilde{\Lambda}_{\textrm{X1n}}^{\textrm{f}}$[meV]}\\
            \colrule
            0.54 & 0.02 & 41.0 & -7.5 & 1.4 & 8.4\\
        \end{tabular}
    \end{ruledtabular}
\end{table}
%-----------------------------------------------

%-----------------------------------------------
%-----------------------------------------------

Compared to the top configuration, all spin-orbit parameters are of the spin-flipping nature (superscript f on $\Lambda$s). Spin-conserving hoppings are also allowed by the local symmetry but, as we tested, they were not important to fit the DFT data. Hence we did not include them in the minimal SOC Hamiltonian $\mathcal{H}_{\rm soc}$ represented by Eq.~(\ref{Eq:HSOB}).
When comparing the top and bridge configurations we see that the SOC strengths are ranging between $10-50\,\mathrm{meV}$ and $1-40$\,meV, respectively. We note that an experimental prediction of 20~meV\cite{Balakrishnan2014} lies in the range of our parameters.

%-----------------------------------------------
\section{\label{sec:single_adatom}Single adatom limit - resonant states}
%-----------------------------------------------

%-----------------------------------------------
\begin{figure}
	\begin{minipage}{0.49\columnwidth}
 	\includegraphics[width = \columnwidth]{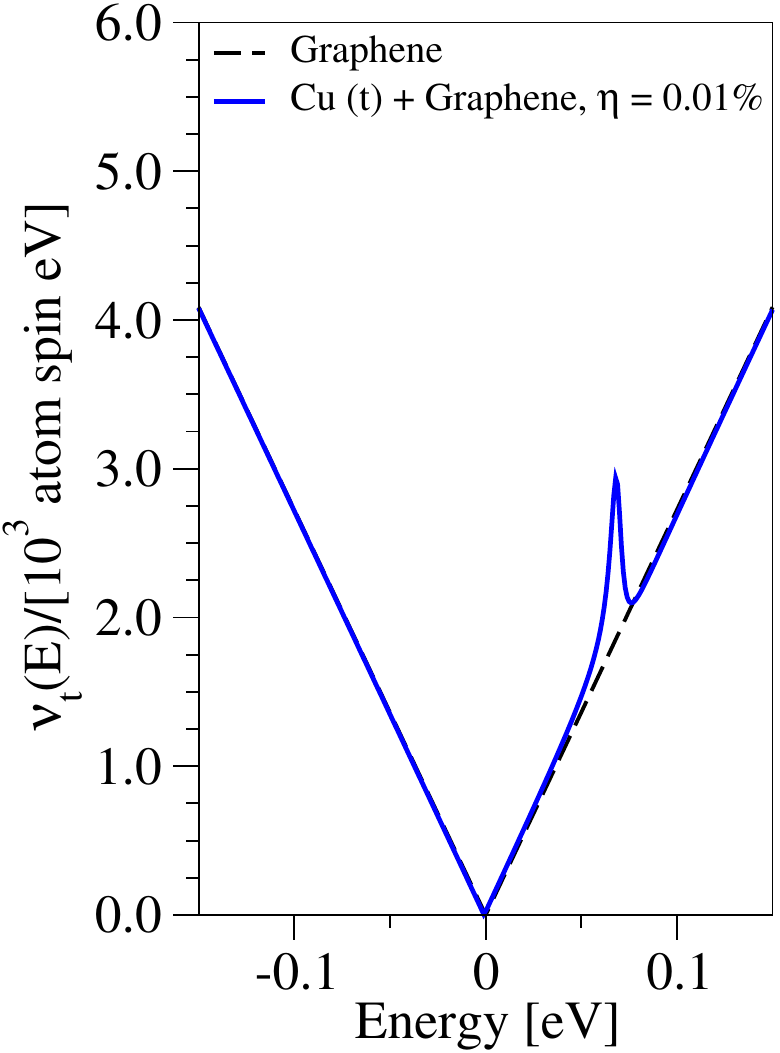}
 	\end{minipage}
 	\begin{minipage}{0.49\columnwidth}
 	\includegraphics[width = \columnwidth]{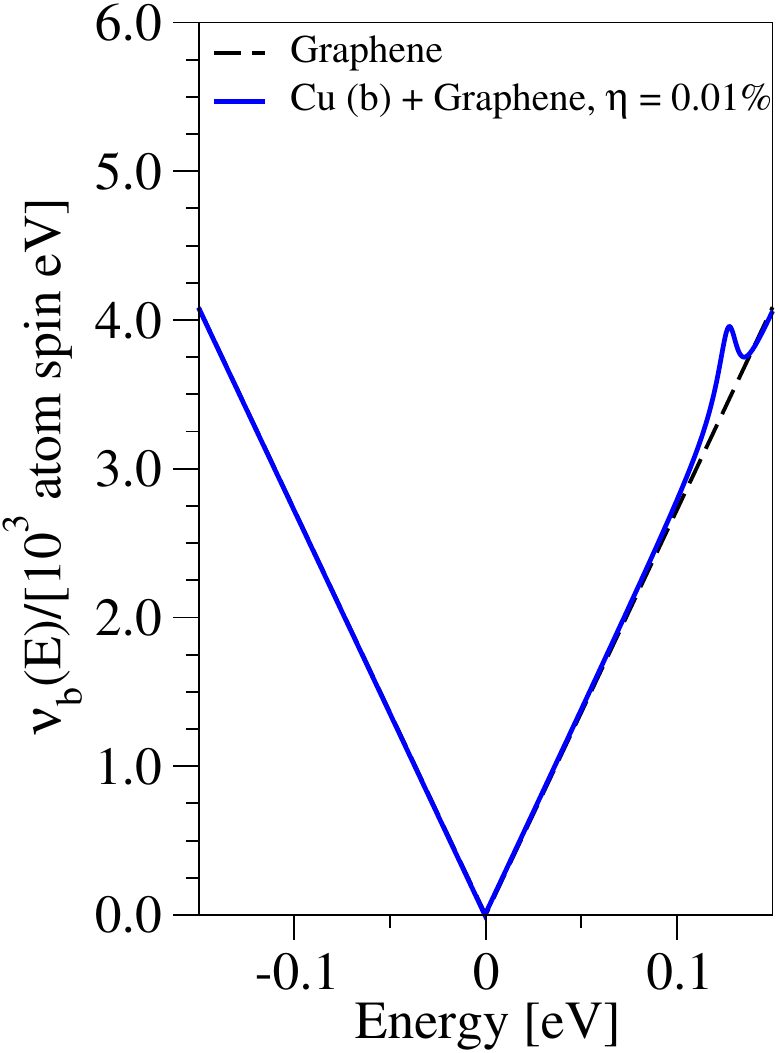}
 	\end{minipage}
 	\caption{\label{fig:pertdos_tb} (Color online) Perturbed DOS $\nu_p$ of copper binding in (a) top, $p=t$, and (b) bridge position, $p=b$, for impurity concentration $\eta = 0.01\%$.}
\end{figure}
%-----------------------------------------------

Ignoring the SOC terms in the full model Hamiltonian $\mathcal{H}=\mathcal{H}_0+\mathcal{H}_{\rm orb}+\mathcal{H}_{\rm soc}$, and assuming that the fitted orbital parameters $\omega$ and $\varepsilon$ are representative enough to describe the single adatom limit (infinite supercell), we now discuss the resonant properties of copper in the top and bridge positions. The general strategy is to eliminate the copper degrees of freedom in $\mathcal{H}$ and then investigate the reduced system (pristine graphene + on-site perturbations) by means of the T-matrix\cite{Hewson1993} and Green's functions of the unperturbed graphene\cite{Ducastelle2013:PRB,Bundesmann2015} treated in position space:
\begin{equation}
G_{c_i,c_j}(E)=\langle c_i|(E+\ii \delta -\mathcal{H}_0)^{-1}|c_j\rangle\,.
\end{equation}
Here, $\delta$ is an infinitesimal positive quantity and $|c_i\rangle$ represents the carbon $p_z$-orbital at lattice site $i$. Since the SOC term is suppressed in $\mathcal{H}_0$, the Green's function elements are spin independent (and that is why we discarded the spin indices in the corresponding $|c_i\rangle$s).

The two pristine graphene Green's functions we need are the on-site function $G_0(E)=G_{c_i,c_i}(E)$, which is independent of the lattice site:
\begin{equation}
G_0(E)\simeq\frac{E}{D^2}\Bigl[\ln{\Bigl|\frac{E^2}{D^2-E^2}\Bigr|}-\ii\,\pi\,\mathrm{sgn}(E)\,\Theta(D-|E|)\Bigr]\,,
\end{equation}
and the function $G_{AB}(E)=\langle C_{\rm A}|(E+\ii \delta -\mathcal{H}_0)^{-1}|C_{\rm B}\rangle$ that couples two particular neighboring sites
$\rm{C_A}$ and $\rm{C_B}$ displayed in Fig.~\ref{fig:geometries}(b):
\begin{align}
G_{\rm A,B}(E)\simeq &\frac{3t}{D^2}\Bigl[1+\ii\,\pi\,\mathrm{sgn}(E)\,\frac{E^2}{9 t^2}\Bigr]\,.
\end{align}
In the above formulas, $D=\sqrt{\sqrt{3}\pi}t$ stands for the effective graphene bandwidth, $t=2.6$\,eV is the standard nearest neighbor hopping, and $\Theta$
is the Heaviside step function.

Starting with the orbital Hamiltonian $\mathcal{H}$ for copper in the top position, and downfolding the $|X\rangle$ orbital by means of the L\"{o}wdin transformation, we arrive at the Hamiltonian $\mathcal{H}_0$ (without SOC term) + $\mathcal{H}'_t$. The on-site perturbation $\mathcal{H}'_t$ is given as follows,
\begin{align}
\mathcal{H}'_t\left(E\right) =  \alpha(E)\sum\limits_\sigma |C_\sigma\rangle \langle C_\sigma|,
\end{align}
where
\begin{equation}
\alpha(E) = \frac{\left(\omega_t\right)^2}{E-\varepsilon_t},
\end{equation}
and $|C\rangle$ is the $p_z$-orbital on the functionalized carbon atom displayed in Fig.~\ref{fig:geometries}(a).

Starting with the full Hamiltonian $\mathcal{H}$ for the bridge position, ignoring SOC terms and downfolding the copper orbital as before, we arrive at the Hamiltonian $\mathcal{H}_0$ (without SOC term) + $\mathcal{H}'_b$, where the on-site perturbation is
\begin{align}
\mathcal{H}'_b\left(E\right) &=\beta(E)\sum\limits_\sigma \left(|C_{\mathrm{A},\s}\rangle+|C_{\mathrm{B},\s}\rangle\right)\left(\langle C_{\mathrm{A},\s}|+\langle C_{\mathrm{B},\s}|\right)\,.
\end{align}
Similarly as before,
\begin{equation}
\beta(E) = \frac{\left(\omega_b\right)^2}{E-\varepsilon_b},
\end{equation}
and $|C_{\mathrm{A}}\rangle$ and $|C_{\mathrm{B}}\rangle$ are carbon $p_z$-orbitals (again with the suppressed spin index) on the functionalized carbon sites
$\rm{C_A}$ and $\rm{C_B}$, respectively, see Fig.~\ref{fig:geometries}(b).

For both cases, the T-matrix in the presence of perturbation $\mathcal{H}'$ reads,
\begin{equation}
\mathrm{T}\left(E\right) = \left[1-\mathcal{H}'\left(E+\ii\delta-\mathcal{H}_0\right)^{-1}\right]^{-1}\,\mathcal{H}'\,,
\end{equation}
and consequently the change in the DOS, $\Delta \nu\left(E\right)$, becomes,
\begin{align}\label{Eq:pertdos}
\Delta \nu\left(E\right) = \frac{1}{\pi} \mathrm{Im} \left\{\mathrm{Tr}\left[\mathrm{T}\left(E\right)\frac{\partial}{\partial E}\left(E+\ii\delta-\mathcal{H}_0\right)^{-1} \right] \right\}\,.
\end{align}
Particularly, for the top adsorbed impurity Eq.~(\ref{Eq:pertdos}) gives,
\begin{equation}\label{Eq:pertdos_top}
\Delta\nu_t(E)=\frac{1}{\pi}\,\mathrm{Im}\Bigl[\frac{\alpha(E)}{1-\alpha(E)\,G_0(E)}\,\frac{\partial}{\partial E}\, G_0(E)\Bigr]\,,
\end{equation}
while for the bridge one we arrive at,
\begin{align}\label{Eq:pertdos_bridge}
\Delta \nu_b\left(E\right) = \frac{1}{\pi}\,\mathrm{Im}&\Bigl[\frac{2\beta(E)}{1-2\beta(E)\left[G_0(E)+G_{\rm A,B}(E)\right]}
\nonumber\\
&\times\left(\frac{\partial}{\partial E}\, G_{0}(E) + \frac{\partial}{\partial E}\, G_{\rm A,B}(E)\right)\Bigr]\,.
\end{align}

Figures~\ref{fig:pertdos_tb}(a) and (b) display the resulting perturbed DOS per atom and spin, $\nu(E) = \nu_0(E)+\eta\Delta\nu(E)$, for the top and bridge adsorption position, respectively. Here $\nu_0(E)=|E|/D^2$ is the unperturbed graphene density per atom and spin, $\Delta\nu(E)$ is given either by Eq.~(\ref{Eq:pertdos_top}) or by (\ref{Eq:pertdos_bridge}), depending on the adsorption position, and $\eta$ stands for the concentration of copper per number of carbon atoms. We observe that the resonance levels develop at energies $E^{t}_\mathrm{res} \simeq 69\,\mathrm{meV}$ and $E^{b}_\mathrm{res} \simeq 128\,\mathrm{meV}$ with the corresponding widths $\Gamma^t\sim 6.9\,\mathrm{meV}$ and $\Gamma^b\sim 12\,\mathrm{meV}$.

Both resonance peaks are very narrow and relatively close to the graphene charge neutrality point---the top position resonance is by one half closer than the bridge one. Therefore copper is expected to behave as a resonant scatterer in the limit of small impurity concentrations.
The lifetimes of the resonances, $\tau \sim \hbar/\Gamma$, are quite large: $\tau_t\simeq 100$\,fs and $\tau_b\simeq 50$\,fs, respectively. Since the copper atom induces strong local spin-orbit coupling---$\Lambda$s are larger than the resonance widths $\Gamma$s---and possesses also a large magnetic moment we expect that it acts as a spin hot spot in spin relaxation. \cite{Kochan2014, Kochan2015}

%-----------------------------------------------
\section{\label{sec:concl} Summary}
%-----------------------------------------------
We performed systematic DFT and phenomenological investigations of copper adsorbed to graphene on the top and bridge positions. Although bridge and top positions of copper on graphene are energetically close in binding energies, they significantly differ in the local symmetry. However, the orbital physics in both systems is remarkably similar, namely $p$ and $d$ orbitals contribute at the Fermi energy. The main mechanism for the induced spin-orbit coupling is the presence of these Cu $p$ and $d$ orbitals, which are equally important. To quantify spin-orbit coupling further, we constructed model Hamiltonians, which can be used to fit the low-energy ab-initio data over the whole Brillouin zone and in this way extract orbital and spin-orbital parameters; the latter are ranging in the tens of meV. By application of the extracted single adatom model to the Green's function and T-matrix formalism we showed that copper atoms act as resonant scatterers.

Our study should motivate further experimental investigations of the spin Hall effect in copper decorated graphene. We expect that our models reliably describe the low-energy physics of copper on graphene, allowing for transport and spin relaxation simulations at large-scale.\cite{Tuan2014}

\begin{acknowledgments}
    This work was supported by the DFG SFB Grant No. 689 and GRK Grant No. 1570, and by the EU Seventh Framework Programme under Grant Agreement No. 604391 Graphene Flagship. The authors gratefully acknowledge the Gauss Centre for Supercomputing e.V. (www.gauss-centre.eu) for funding this project by providing computing time on the GCS Supercomputer SuperMUC at Leibniz Supercomputing Centre (LRZ, www.lrz.de).
\end{acknowledgments}

\bibliography{paper}

\begin{thebibliography}{44}%
\makeatletter
\providecommand \@ifxundefined [1]{%
 \@ifx{#1\undefined}
}%
\providecommand \@ifnum [1]{%
 \ifnum #1\expandafter \@firstoftwo
 \else \expandafter \@secondoftwo
 \fi
}%
\providecommand \@ifx [1]{%
 \ifx #1\expandafter \@firstoftwo
 \else \expandafter \@secondoftwo
 \fi
}%
\providecommand \natexlab [1]{#1}%
\providecommand \enquote  [1]{``#1''}%
\providecommand \bibnamefont  [1]{#1}%
\providecommand \bibfnamefont [1]{#1}%
\providecommand \citenamefont [1]{#1}%
\providecommand \href@noop [0]{\@secondoftwo}%
\providecommand \href [0]{\begingroup \@sanitize@url \@href}%
\providecommand \@href[1]{\@@startlink{#1}\@@href}%
\providecommand \@@href[1]{\endgroup#1\@@endlink}%
\providecommand \@sanitize@url [0]{\catcode `\\12\catcode `\$12\catcode
  `\&12\catcode `\#12\catcode `\^12\catcode `\_12\catcode `\%12\relax}%
\providecommand \@@startlink[1]{}%
\providecommand \@@endlink[0]{}%
\providecommand \url  [0]{\begingroup\@sanitize@url \@url }%
\providecommand \@url [1]{\endgroup\@href {#1}{\urlprefix }}%
\providecommand \urlprefix  [0]{URL }%
\providecommand \Eprint [0]{\href }%
\providecommand \doibase [0]{http://dx.doi.org/}%
\providecommand \selectlanguage [0]{\@gobble}%
\providecommand \bibinfo  [0]{\@secondoftwo}%
\providecommand \bibfield  [0]{\@secondoftwo}%
\providecommand \translation [1]{[#1]}%
\providecommand \BibitemOpen [0]{}%
\providecommand \bibitemStop [0]{}%
\providecommand \bibitemNoStop [0]{.\EOS\space}%
\providecommand \EOS [0]{\spacefactor3000\relax}%
\providecommand \BibitemShut  [1]{\csname bibitem#1\endcsname}%
\let\auto@bib@innerbib\@empty
%</preamble>
\bibitem [{\citenamefont {{Castro Neto}}\ and\ \citenamefont
  {Guinea}(2009)}]{Neto2009}%
  \BibitemOpen
  \bibfield  {author} {\bibinfo {author} {\bibfnamefont {A.~H.}\ \bibnamefont
  {{Castro Neto}}}\ and\ \bibinfo {author} {\bibfnamefont {F.}~\bibnamefont
  {Guinea}},\ }\href {http://link.aps.org/doi/10.1103/PhysRevLett.103.026804}
  {\bibfield  {journal} {\bibinfo  {journal} {Phys. Rev. Lett.}\ }\textbf
  {\bibinfo {volume} {103}},\ \bibinfo {pages} {026804} (\bibinfo {year}
  {2009})}\BibitemShut {NoStop}%
\bibitem [{\citenamefont {Han}\ \emph {et~al.}(2014)\citenamefont {Han},
  \citenamefont {Kawakami}, \citenamefont {Gmitra},\ and\ \citenamefont
  {Fabian}}]{Han2014}%
  \BibitemOpen
  \bibfield  {author} {\bibinfo {author} {\bibfnamefont {W.}~\bibnamefont
  {Han}}, \bibinfo {author} {\bibfnamefont {R.~K.}\ \bibnamefont {Kawakami}},
  \bibinfo {author} {\bibfnamefont {M.}~\bibnamefont {Gmitra}}, \ and\ \bibinfo
  {author} {\bibfnamefont {J.}~\bibnamefont {Fabian}},\ }\href {\doibase
  10.1038/nnano.2014.214} {\bibfield  {journal} {\bibinfo  {journal} {Nat.
  Nano}\ }\textbf {\bibinfo {volume} {9}},\ \bibinfo {pages} {794} (\bibinfo
  {year} {2014})}\BibitemShut {NoStop}%
\bibitem [{\citenamefont {{\v{Z}}uti{\'{c}}}\ \emph {et~al.}(2004)\citenamefont
  {{\v{Z}}uti{\'{c}}}, \citenamefont {Fabian},\ and\ \citenamefont
  {Sarma}}]{Zutic2004}%
  \BibitemOpen
  \bibfield  {author} {\bibinfo {author} {\bibfnamefont {I.}~\bibnamefont
  {{\v{Z}}uti{\'{c}}}}, \bibinfo {author} {\bibfnamefont {J.}~\bibnamefont
  {Fabian}}, \ and\ \bibinfo {author} {\bibfnamefont {S.~D.}\ \bibnamefont
  {Sarma}},\ }\href {\doibase 10.1103/RevModPhys.76.323} {\bibfield  {journal}
  {\bibinfo  {journal} {Rev. Mod. Phys.}\ }\textbf {\bibinfo {volume} {76}},\
  \bibinfo {pages} {323} (\bibinfo {year} {2004})}\BibitemShut {NoStop}%
\bibitem [{\citenamefont {Fabian}(2007)}]{Fabian2007}%
  \BibitemOpen
  \bibfield  {author} {\bibinfo {author} {\bibfnamefont {J.}~\bibnamefont
  {Fabian}},\ }\href@noop {} {\bibfield  {journal} {\bibinfo  {journal} {Acta
  Phys. Slovaca}\ }\textbf {\bibinfo {volume} {57}},\ \bibinfo {pages} {565}
  (\bibinfo {year} {2007})}\BibitemShut {NoStop}%
\bibitem [{\citenamefont {Han}\ and\ \citenamefont {Kawakami}(2011)}]{Han2011}%
  \BibitemOpen
  \bibfield  {author} {\bibinfo {author} {\bibfnamefont {W.}~\bibnamefont
  {Han}}\ and\ \bibinfo {author} {\bibfnamefont {R.~K.}\ \bibnamefont
  {Kawakami}},\ }\href@noop {} {\bibfield  {journal} {\bibinfo  {journal}
  {Phys. Rev. Lett.}\ }\textbf {\bibinfo {volume} {107}},\ \bibinfo {pages}
  {047207} (\bibinfo {year} {2011})}\BibitemShut {NoStop}%
\bibitem [{\citenamefont {Balakrishnan}\ \emph {et~al.}(2013)\citenamefont
  {Balakrishnan}, \citenamefont {{Kok Wai Koon}}, \citenamefont {Jaiswal},
  \citenamefont {{Castro Neto}},\ and\ \citenamefont
  {{\"{O}}zyilmaz}}]{Balakrishnan2013}%
  \BibitemOpen
  \bibfield  {author} {\bibinfo {author} {\bibfnamefont {J.}~\bibnamefont
  {Balakrishnan}}, \bibinfo {author} {\bibfnamefont {G.}~\bibnamefont {{Kok Wai
  Koon}}}, \bibinfo {author} {\bibfnamefont {M.}~\bibnamefont {Jaiswal}},
  \bibinfo {author} {\bibfnamefont {A.~H.}\ \bibnamefont {{Castro Neto}}}, \
  and\ \bibinfo {author} {\bibfnamefont {B.}~\bibnamefont {{\"{O}}zyilmaz}},\
  }\href {http://www.nature.com/doifinder/10.1038/nphys2576} {\bibfield
  {journal} {\bibinfo  {journal} {Nat. Phys.}\ }\textbf {\bibinfo {volume}
  {9}},\ \bibinfo {pages} {284} (\bibinfo {year} {2013})}\BibitemShut {NoStop}%
\bibitem [{\citenamefont {Gonz{\'{a}}lez-Herrero}\ \emph
  {et~al.}(2016)\citenamefont {Gonz{\'{a}}lez-Herrero}, \citenamefont
  {G{\'{o}}mez-Rodr{\'{i}}guez}, \citenamefont {Mallet}, \citenamefont
  {Moaied}, \citenamefont {Palacios}, \citenamefont {Salgado}, \citenamefont
  {Ugeda}, \citenamefont {Veuillen}, \citenamefont {Yndurain},\ and\
  \citenamefont {Brihuega}}]{Gonzalez-Herrero2016}%
  \BibitemOpen
  \bibfield  {author} {\bibinfo {author} {\bibfnamefont {H.}~\bibnamefont
  {Gonz{\'{a}}lez-Herrero}}, \bibinfo {author} {\bibfnamefont {J.~M.}\
  \bibnamefont {G{\'{o}}mez-Rodr{\'{i}}guez}}, \bibinfo {author} {\bibfnamefont
  {P.}~\bibnamefont {Mallet}}, \bibinfo {author} {\bibfnamefont
  {M.}~\bibnamefont {Moaied}}, \bibinfo {author} {\bibfnamefont {J.~J.}\
  \bibnamefont {Palacios}}, \bibinfo {author} {\bibfnamefont {C.}~\bibnamefont
  {Salgado}}, \bibinfo {author} {\bibfnamefont {M.~M.}\ \bibnamefont {Ugeda}},
  \bibinfo {author} {\bibfnamefont {J.-Y.}\ \bibnamefont {Veuillen}}, \bibinfo
  {author} {\bibfnamefont {F.}~\bibnamefont {Yndurain}}, \ and\ \bibinfo
  {author} {\bibfnamefont {I.}~\bibnamefont {Brihuega}},\ }\href {\doibase
  10.1126/science.aad8038} {\bibfield  {journal} {\bibinfo  {journal}
  {Science}\ }\textbf {\bibinfo {volume} {352}},\ \bibinfo {pages} {437}
  (\bibinfo {year} {2016})}\BibitemShut {NoStop}%
\bibitem [{\citenamefont {Yazyev}\ and\ \citenamefont
  {Helm}(2007)}]{Yazyev2007}%
  \BibitemOpen
  \bibfield  {author} {\bibinfo {author} {\bibfnamefont {O.~V.}\ \bibnamefont
  {Yazyev}}\ and\ \bibinfo {author} {\bibfnamefont {L.}~\bibnamefont {Helm}},\
  }\href {http://link.aps.org/doi/10.1103/PhysRevB.75.125408} {\bibfield
  {journal} {\bibinfo  {journal} {Phys. Rev. B}\ }\textbf {\bibinfo {volume}
  {75}},\ \bibinfo {pages} {125408} (\bibinfo {year} {2007})}\BibitemShut
  {NoStop}%
\bibitem [{\citenamefont {Gmitra}\ \emph {et~al.}(2013)\citenamefont {Gmitra},
  \citenamefont {Kochan},\ and\ \citenamefont {Fabian}}]{Gmitra2013}%
  \BibitemOpen
  \bibfield  {author} {\bibinfo {author} {\bibfnamefont {M.}~\bibnamefont
  {Gmitra}}, \bibinfo {author} {\bibfnamefont {D.}~\bibnamefont {Kochan}}, \
  and\ \bibinfo {author} {\bibfnamefont {J.}~\bibnamefont {Fabian}},\ }\href
  {\doibase 10.1103/PhysRevLett.110.246602} {\bibfield  {journal} {\bibinfo
  {journal} {Phys. Rev. Lett.}\ }\textbf {\bibinfo {volume} {110}},\ \bibinfo
  {pages} {246602} (\bibinfo {year} {2013})}\BibitemShut {NoStop}%
\bibitem [{\citenamefont {Kochan}\ \emph {et~al.}(2014)\citenamefont {Kochan},
  \citenamefont {Gmitra},\ and\ \citenamefont {Fabian}}]{Kochan2014}%
  \BibitemOpen
  \bibfield  {author} {\bibinfo {author} {\bibfnamefont {D.}~\bibnamefont
  {Kochan}}, \bibinfo {author} {\bibfnamefont {M.}~\bibnamefont {Gmitra}}, \
  and\ \bibinfo {author} {\bibfnamefont {J.}~\bibnamefont {Fabian}},\ }\href
  {\doibase 10.1103/PhysRevLett.112.116602} {\bibfield  {journal} {\bibinfo
  {journal} {Phys. Rev. Lett.}\ }\textbf {\bibinfo {volume} {112}},\ \bibinfo
  {pages} {116602} (\bibinfo {year} {2014})}\BibitemShut {NoStop}%
\bibitem [{\citenamefont {Gmitra}\ \emph {et~al.}(2009)\citenamefont {Gmitra},
  \citenamefont {Konschuh}, \citenamefont {Ertler}, \citenamefont
  {Ambrosch-Draxl},\ and\ \citenamefont {Fabian}}]{Gmitra2009}%
  \BibitemOpen
  \bibfield  {author} {\bibinfo {author} {\bibfnamefont {M.}~\bibnamefont
  {Gmitra}}, \bibinfo {author} {\bibfnamefont {S.}~\bibnamefont {Konschuh}},
  \bibinfo {author} {\bibfnamefont {C.}~\bibnamefont {Ertler}}, \bibinfo
  {author} {\bibfnamefont {C.}~\bibnamefont {Ambrosch-Draxl}}, \ and\ \bibinfo
  {author} {\bibfnamefont {J.}~\bibnamefont {Fabian}},\ }\href {\doibase
  10.1103/PhysRevB.80.235431} {\bibfield  {journal} {\bibinfo  {journal} {Phys.
  Rev. B}\ }\textbf {\bibinfo {volume} {80}},\ \bibinfo {pages} {235431}
  (\bibinfo {year} {2009})}\BibitemShut {NoStop}%
\bibitem [{\citenamefont {Zollner}\ \emph {et~al.}(2016)\citenamefont
  {Zollner}, \citenamefont {Frank}, \citenamefont {Irmer}, \citenamefont
  {Gmitra}, \citenamefont {Kochan},\ and\ \citenamefont
  {Fabian}}]{Zollner2016}%
  \BibitemOpen
  \bibfield  {author} {\bibinfo {author} {\bibfnamefont {K.}~\bibnamefont
  {Zollner}}, \bibinfo {author} {\bibfnamefont {T.}~\bibnamefont {Frank}},
  \bibinfo {author} {\bibfnamefont {S.}~\bibnamefont {Irmer}}, \bibinfo
  {author} {\bibfnamefont {M.}~\bibnamefont {Gmitra}}, \bibinfo {author}
  {\bibfnamefont {D.}~\bibnamefont {Kochan}}, \ and\ \bibinfo {author}
  {\bibfnamefont {J.}~\bibnamefont {Fabian}},\ }\href {\doibase
  10.1103/PhysRevB.93.045423} {\bibfield  {journal} {\bibinfo  {journal} {Phys.
  Rev. B}\ }\textbf {\bibinfo {volume} {93}},\ \bibinfo {pages} {045423}
  (\bibinfo {year} {2016})}\BibitemShut {NoStop}%
\bibitem [{\citenamefont {Irmer}\ \emph {et~al.}(2015)\citenamefont {Irmer},
  \citenamefont {Frank}, \citenamefont {Putz}, \citenamefont {Gmitra},
  \citenamefont {Kochan},\ and\ \citenamefont {Fabian}}]{Irmer2015}%
  \BibitemOpen
  \bibfield  {author} {\bibinfo {author} {\bibfnamefont {S.}~\bibnamefont
  {Irmer}}, \bibinfo {author} {\bibfnamefont {T.}~\bibnamefont {Frank}},
  \bibinfo {author} {\bibfnamefont {S.}~\bibnamefont {Putz}}, \bibinfo {author}
  {\bibfnamefont {M.}~\bibnamefont {Gmitra}}, \bibinfo {author} {\bibfnamefont
  {D.}~\bibnamefont {Kochan}}, \ and\ \bibinfo {author} {\bibfnamefont
  {J.}~\bibnamefont {Fabian}},\ }\href {\doibase 10.1103/PhysRevB.91.115141}
  {\bibfield  {journal} {\bibinfo  {journal} {Phys. Rev. B}\ }\textbf {\bibinfo
  {volume} {91}},\ \bibinfo {pages} {115141} (\bibinfo {year}
  {2015})}\BibitemShut {NoStop}%
\bibitem [{\citenamefont {Guzm\'an-Arellano}\ \emph {et~al.}(2015)\citenamefont
  {Guzm\'an-Arellano}, \citenamefont {Hern\'andez-Nieves}, \citenamefont
  {Balseiro},\ and\ \citenamefont {Usaj}}]{Guzman2015}%
  \BibitemOpen
  \bibfield  {author} {\bibinfo {author} {\bibfnamefont {R.~M.}\ \bibnamefont
  {Guzm\'an-Arellano}}, \bibinfo {author} {\bibfnamefont {A.~D.}\ \bibnamefont
  {Hern\'andez-Nieves}}, \bibinfo {author} {\bibfnamefont {C.~A.}\ \bibnamefont
  {Balseiro}}, \ and\ \bibinfo {author} {\bibfnamefont {G.}~\bibnamefont
  {Usaj}},\ }\href {\doibase 10.1103/PhysRevB.91.195408} {\bibfield  {journal}
  {\bibinfo  {journal} {Phys. Rev. B}\ }\textbf {\bibinfo {volume} {91}},\
  \bibinfo {pages} {195408} (\bibinfo {year} {2015})}\BibitemShut {NoStop}%
\bibitem [{\citenamefont {Hu}\ \emph {et~al.}(2012)\citenamefont {Hu},
  \citenamefont {Alicea}, \citenamefont {Wu},\ and\ \citenamefont
  {Franz}}]{Hu2012}%
  \BibitemOpen
  \bibfield  {author} {\bibinfo {author} {\bibfnamefont {J.}~\bibnamefont
  {Hu}}, \bibinfo {author} {\bibfnamefont {J.}~\bibnamefont {Alicea}}, \bibinfo
  {author} {\bibfnamefont {R.}~\bibnamefont {Wu}}, \ and\ \bibinfo {author}
  {\bibfnamefont {M.}~\bibnamefont {Franz}},\ }\href {\doibase
  10.1103/PhysRevLett.109.266801} {\bibfield  {journal} {\bibinfo  {journal}
  {Phys. Rev. Lett.}\ }\textbf {\bibinfo {volume} {109}},\ \bibinfo {pages}
  {266801} (\bibinfo {year} {2012})}\BibitemShut {NoStop}%
\bibitem [{\citenamefont {Ma}\ \emph {et~al.}(2012)\citenamefont {Ma},
  \citenamefont {Li},\ and\ \citenamefont {Yang}}]{Ma2012}%
  \BibitemOpen
  \bibfield  {author} {\bibinfo {author} {\bibfnamefont {D.}~\bibnamefont
  {Ma}}, \bibinfo {author} {\bibfnamefont {Z.}~\bibnamefont {Li}}, \ and\
  \bibinfo {author} {\bibfnamefont {Z.}~\bibnamefont {Yang}},\ }\href
  {http://dx.doi.org/10.1016/j.carbon.2011.08.055} {\bibfield  {journal}
  {\bibinfo  {journal} {Carbon}\ }\textbf {\bibinfo {volume} {50}},\ \bibinfo
  {pages} {297} (\bibinfo {year} {2012})}\BibitemShut {NoStop}%
\bibitem [{\citenamefont {Weeks}\ \emph {et~al.}(2011)\citenamefont {Weeks},
  \citenamefont {Hu}, \citenamefont {Alicea}, \citenamefont {Franz},\ and\
  \citenamefont {Wu}}]{Weeks2011}%
  \BibitemOpen
  \bibfield  {author} {\bibinfo {author} {\bibfnamefont {C.}~\bibnamefont
  {Weeks}}, \bibinfo {author} {\bibfnamefont {J.}~\bibnamefont {Hu}}, \bibinfo
  {author} {\bibfnamefont {J.}~\bibnamefont {Alicea}}, \bibinfo {author}
  {\bibfnamefont {M.}~\bibnamefont {Franz}}, \ and\ \bibinfo {author}
  {\bibfnamefont {R.}~\bibnamefont {Wu}},\ }\href {\doibase
  10.1103/PhysRevX.1.021001} {\bibfield  {journal} {\bibinfo  {journal} {Phys.
  Rev. X}\ }\textbf {\bibinfo {volume} {1}},\ \bibinfo {pages} {021001}
  (\bibinfo {year} {2011})}\BibitemShut {NoStop}%
\bibitem [{\citenamefont {Wehling}\ \emph {et~al.}(2010)\citenamefont
  {Wehling}, \citenamefont {Yuan}, \citenamefont {Lichtenstein}, \citenamefont
  {Geim},\ and\ \citenamefont {Katsnelson}}]{Wehling2010}%
  \BibitemOpen
  \bibfield  {author} {\bibinfo {author} {\bibfnamefont {T.~O.}\ \bibnamefont
  {Wehling}}, \bibinfo {author} {\bibfnamefont {S.}~\bibnamefont {Yuan}},
  \bibinfo {author} {\bibfnamefont {A.~I.}\ \bibnamefont {Lichtenstein}},
  \bibinfo {author} {\bibfnamefont {A.~K.}\ \bibnamefont {Geim}}, \ and\
  \bibinfo {author} {\bibfnamefont {M.~I.}\ \bibnamefont {Katsnelson}},\ }\href
  {\doibase 10.1103/PhysRevLett.105.056802} {\bibfield  {journal} {\bibinfo
  {journal} {Phys. Rev. Lett.}\ }\textbf {\bibinfo {volume} {105}},\ \bibinfo
  {pages} {056802} (\bibinfo {year} {2010})}\BibitemShut {NoStop}%
\bibitem [{\citenamefont {Hong}\ \emph {et~al.}(2011)\citenamefont {Hong},
  \citenamefont {Cheng}, \citenamefont {Herding},\ and\ \citenamefont
  {Zhu}}]{Hong2011}%
  \BibitemOpen
  \bibfield  {author} {\bibinfo {author} {\bibfnamefont {X.}~\bibnamefont
  {Hong}}, \bibinfo {author} {\bibfnamefont {S.-H.}\ \bibnamefont {Cheng}},
  \bibinfo {author} {\bibfnamefont {C.}~\bibnamefont {Herding}}, \ and\
  \bibinfo {author} {\bibfnamefont {J.}~\bibnamefont {Zhu}},\ }\href {\doibase
  10.1103/PhysRevB.83.085410} {\bibfield  {journal} {\bibinfo  {journal} {Phys.
  Rev. B}\ }\textbf {\bibinfo {volume} {83}},\ \bibinfo {pages} {085410}
  (\bibinfo {year} {2011})}\BibitemShut {NoStop}%
\bibitem [{\citenamefont {Ferreira}\ \emph {et~al.}(2014)\citenamefont
  {Ferreira}, \citenamefont {Rappoport}, \citenamefont {Cazalilla},\ and\
  \citenamefont {{Castro Neto}}}]{Ferreira2014}%
  \BibitemOpen
  \bibfield  {author} {\bibinfo {author} {\bibfnamefont {A.}~\bibnamefont
  {Ferreira}}, \bibinfo {author} {\bibfnamefont {T.~G.}\ \bibnamefont
  {Rappoport}}, \bibinfo {author} {\bibfnamefont {M.~A.}\ \bibnamefont
  {Cazalilla}}, \ and\ \bibinfo {author} {\bibfnamefont {A.~H.}\ \bibnamefont
  {{Castro Neto}}},\ }\href@noop {} {\bibfield  {journal} {\bibinfo  {journal}
  {Phys. Rev. Lett.}\ }\textbf {\bibinfo {volume} {112}},\ \bibinfo {pages}
  {066601} (\bibinfo {year} {2014})}\BibitemShut {NoStop}%
\bibitem [{\citenamefont {Tuan}\ \emph {et~al.}(2014)\citenamefont {Tuan},
  \citenamefont {Ortmann}, \citenamefont {Soriano}, \citenamefont
  {Valenzuela},\ and\ \citenamefont {Roche}}]{Tuan2014}%
  \BibitemOpen
  \bibfield  {author} {\bibinfo {author} {\bibfnamefont {D.~V.}\ \bibnamefont
  {Tuan}}, \bibinfo {author} {\bibfnamefont {F.}~\bibnamefont {Ortmann}},
  \bibinfo {author} {\bibfnamefont {D.}~\bibnamefont {Soriano}}, \bibinfo
  {author} {\bibfnamefont {S.~O.}\ \bibnamefont {Valenzuela}}, \ and\ \bibinfo
  {author} {\bibfnamefont {S.}~\bibnamefont {Roche}},\ }\href@noop {}
  {\bibfield  {journal} {\bibinfo  {journal} {Nat. Phys.}\ }\textbf {\bibinfo
  {volume} {10}},\ \bibinfo {pages} {857} (\bibinfo {year} {2014})}\BibitemShut
  {NoStop}%
\bibitem [{\citenamefont {Chan}\ \emph {et~al.}(2008)\citenamefont {Chan},
  \citenamefont {Neaton},\ and\ \citenamefont {Cohen}}]{Chan2008}%
  \BibitemOpen
  \bibfield  {author} {\bibinfo {author} {\bibfnamefont {K.~T.}\ \bibnamefont
  {Chan}}, \bibinfo {author} {\bibfnamefont {J.~B.}\ \bibnamefont {Neaton}}, \
  and\ \bibinfo {author} {\bibfnamefont {M.~L.}\ \bibnamefont {Cohen}},\ }\href
  {\doibase 10.1103/PhysRevB.77.235430} {\bibfield  {journal} {\bibinfo
  {journal} {Phys. Rev. B}\ }\textbf {\bibinfo {volume} {77}},\ \bibinfo
  {pages} {235430} (\bibinfo {year} {2008})}\BibitemShut {NoStop}%
\bibitem [{\citenamefont {Mattevi}\ \emph {et~al.}(2011)\citenamefont
  {Mattevi}, \citenamefont {Kim},\ and\ \citenamefont
  {Chhowalla}}]{Mattevi2011}%
  \BibitemOpen
  \bibfield  {author} {\bibinfo {author} {\bibfnamefont {C.}~\bibnamefont
  {Mattevi}}, \bibinfo {author} {\bibfnamefont {H.}~\bibnamefont {Kim}}, \ and\
  \bibinfo {author} {\bibfnamefont {M.}~\bibnamefont {Chhowalla}},\ }\href
  {\doibase 10.1039/C0JM02126A} {\bibfield  {journal} {\bibinfo  {journal} {J.
  Mat. Chem.}\ }\textbf {\bibinfo {volume} {21}},\ \bibinfo {pages} {3324}
  (\bibinfo {year} {2011})}\BibitemShut {NoStop}%
\bibitem [{\citenamefont {Balakrishnan}\ \emph {et~al.}(2014)\citenamefont
  {Balakrishnan}, \citenamefont {Koon}, \citenamefont {Avsar}, \citenamefont
  {Ho}, \citenamefont {Lee}, \citenamefont {Jaiswal}, \citenamefont {Baeck},
  \citenamefont {Ahn}, \citenamefont {Ferreira}, \citenamefont {Cazalilla},
  \citenamefont {{Castro Neto}},\ and\ \citenamefont
  {{\"{O}}zyilmaz}}]{Balakrishnan2014}%
  \BibitemOpen
  \bibfield  {author} {\bibinfo {author} {\bibfnamefont {J.}~\bibnamefont
  {Balakrishnan}}, \bibinfo {author} {\bibfnamefont {G.~K.~W.}\ \bibnamefont
  {Koon}}, \bibinfo {author} {\bibfnamefont {A.}~\bibnamefont {Avsar}},
  \bibinfo {author} {\bibfnamefont {Y.}~\bibnamefont {Ho}}, \bibinfo {author}
  {\bibfnamefont {J.~H.}\ \bibnamefont {Lee}}, \bibinfo {author} {\bibfnamefont
  {M.}~\bibnamefont {Jaiswal}}, \bibinfo {author} {\bibfnamefont {S.-J.}\
  \bibnamefont {Baeck}}, \bibinfo {author} {\bibfnamefont {J.-H.}\ \bibnamefont
  {Ahn}}, \bibinfo {author} {\bibfnamefont {A.}~\bibnamefont {Ferreira}},
  \bibinfo {author} {\bibfnamefont {M.~a.}\ \bibnamefont {Cazalilla}}, \bibinfo
  {author} {\bibfnamefont {A.~H.}\ \bibnamefont {{Castro Neto}}}, \ and\
  \bibinfo {author} {\bibfnamefont {B.}~\bibnamefont {{\"{O}}zyilmaz}},\ }\href
  {\doibase 10.1038/ncomms5748} {\bibfield  {journal} {\bibinfo  {journal}
  {Nat. Commun.}\ }\textbf {\bibinfo {volume} {5}},\ \bibinfo {pages} {4748}
  (\bibinfo {year} {2014})}\BibitemShut {NoStop}%
\bibitem [{\citenamefont {Frank}\ \emph {et~al.}(2016)\citenamefont {Frank},
  \citenamefont {Gmitra},\ and\ \citenamefont {Fabian}}]{Frank2016}%
  \BibitemOpen
  \bibfield  {author} {\bibinfo {author} {\bibfnamefont {T.}~\bibnamefont
  {Frank}}, \bibinfo {author} {\bibfnamefont {M.}~\bibnamefont {Gmitra}}, \
  and\ \bibinfo {author} {\bibfnamefont {J.}~\bibnamefont {Fabian}},\ }\href
  {\doibase 10.1103/PhysRevB.93.155142} {\bibfield  {journal} {\bibinfo
  {journal} {Phys. Rev. B}\ }\textbf {\bibinfo {volume} {93}},\ \bibinfo
  {pages} {155142} (\bibinfo {year} {2016})}\BibitemShut {NoStop}%
\bibitem [{\citenamefont {Amft}\ \emph {et~al.}(2011)\citenamefont {Amft},
  \citenamefont {Leb{\`{e}}gue}, \citenamefont {Eriksson},\ and\ \citenamefont
  {Skorodumova}}]{Amft2011}%
  \BibitemOpen
  \bibfield  {author} {\bibinfo {author} {\bibfnamefont {M.}~\bibnamefont
  {Amft}}, \bibinfo {author} {\bibfnamefont {S.}~\bibnamefont {Leb{\`{e}}gue}},
  \bibinfo {author} {\bibfnamefont {O.}~\bibnamefont {Eriksson}}, \ and\
  \bibinfo {author} {\bibfnamefont {N.~V.}\ \bibnamefont {Skorodumova}},\
  }\href {\doibase 10.1088/0953-8984/23/39/395001} {\bibfield  {journal}
  {\bibinfo  {journal} {J. Phys. Cond. Matter}\ }\textbf {\bibinfo {volume}
  {23}},\ \bibinfo {pages} {395001} (\bibinfo {year} {2011})}\BibitemShut
  {NoStop}%
\bibitem [{\citenamefont {Wu}\ \emph {et~al.}(2009)\citenamefont {Wu},
  \citenamefont {Liu}, \citenamefont {Ge},\ and\ \citenamefont
  {Jiang}}]{Wu2009}%
  \BibitemOpen
  \bibfield  {author} {\bibinfo {author} {\bibfnamefont {M.}~\bibnamefont
  {Wu}}, \bibinfo {author} {\bibfnamefont {E.~Z.}\ \bibnamefont {Liu}},
  \bibinfo {author} {\bibfnamefont {M.~Y.}\ \bibnamefont {Ge}}, \ and\ \bibinfo
  {author} {\bibfnamefont {J.~Z.}\ \bibnamefont {Jiang}},\ }\href {\doibase
  10.1063/1.3097013} {\bibfield  {journal} {\bibinfo  {journal} {Appl. Phys.
  Lett.}\ }\textbf {\bibinfo {volume} {94}},\ \bibinfo {pages} {102505}
  (\bibinfo {year} {2009})}\BibitemShut {NoStop}%
\bibitem [{\citenamefont {Cao}\ \emph {et~al.}(2010)\citenamefont {Cao},
  \citenamefont {Wu}, \citenamefont {Jiang},\ and\ \citenamefont
  {Cheng}}]{Cao2010}%
  \BibitemOpen
  \bibfield  {author} {\bibinfo {author} {\bibfnamefont {C.}~\bibnamefont
  {Cao}}, \bibinfo {author} {\bibfnamefont {M.}~\bibnamefont {Wu}}, \bibinfo
  {author} {\bibfnamefont {J.}~\bibnamefont {Jiang}}, \ and\ \bibinfo {author}
  {\bibfnamefont {H.~P.}\ \bibnamefont {Cheng}},\ }\href {\doibase
  10.1103/PhysRevB.81.205424} {\bibfield  {journal} {\bibinfo  {journal} {Phys.
  Rev. B}\ }\textbf {\bibinfo {volume} {81}},\ \bibinfo {pages} {205424}
  (\bibinfo {year} {2010})}\BibitemShut {NoStop}%
\bibitem [{\citenamefont {Bundesmann}\ \emph {et~al.}(2015)\citenamefont
  {Bundesmann}, \citenamefont {Kochan}, \citenamefont {Tkatschenko},
  \citenamefont {Fabian},\ and\ \citenamefont {Richter}}]{Bundesmann2015}%
  \BibitemOpen
  \bibfield  {author} {\bibinfo {author} {\bibfnamefont {J.}~\bibnamefont
  {Bundesmann}}, \bibinfo {author} {\bibfnamefont {D.}~\bibnamefont {Kochan}},
  \bibinfo {author} {\bibfnamefont {F.}~\bibnamefont {Tkatschenko}}, \bibinfo
  {author} {\bibfnamefont {J.}~\bibnamefont {Fabian}}, \ and\ \bibinfo {author}
  {\bibfnamefont {K.}~\bibnamefont {Richter}},\ }\href
  {http://link.aps.org/doi/10.1103/PhysRevB.92.081403} {\bibfield  {journal}
  {\bibinfo  {journal} {Phys. Rev. B}\ }\textbf {\bibinfo {volume} {92}},\
  \bibinfo {pages} {081403(R)} (\bibinfo {year} {2015})}\BibitemShut {NoStop}%
\bibitem [{\citenamefont {Giannozzi}\ \emph {et~al.}(2009)\citenamefont
  {Giannozzi}, \citenamefont {Baroni}, \citenamefont {Bonini}, \citenamefont
  {Calandra}, \citenamefont {Car}, \citenamefont {Cavazzoni}, \citenamefont
  {Ceresoli}, \citenamefont {Chiarotti}, \citenamefont {Cococcioni},
  \citenamefont {Dabo}, \citenamefont {{Dal Corso}}, \citenamefont
  {de~Gironcoli}, \citenamefont {Fabris}, \citenamefont {Fratesi},
  \citenamefont {Gebauer}, \citenamefont {Gerstmann}, \citenamefont
  {Gougoussis}, \citenamefont {Kokalj}, \citenamefont {Lazzeri}, \citenamefont
  {Martin-Samos}, \citenamefont {Marzari}, \citenamefont {Mauri}, \citenamefont
  {Mazzarello}, \citenamefont {Paolini}, \citenamefont {Pasquarello},
  \citenamefont {Paulatto}, \citenamefont {Sbraccia}, \citenamefont {Scandolo},
  \citenamefont {Sclauzero}, \citenamefont {Seitsonen}, \citenamefont
  {Smogunov}, \citenamefont {Umari},\ and\ \citenamefont
  {Wentzcovitch}}]{Giannozzi2009}%
  \BibitemOpen
  \bibfield  {author} {\bibinfo {author} {\bibfnamefont {P.}~\bibnamefont
  {Giannozzi}}, \bibinfo {author} {\bibfnamefont {S.}~\bibnamefont {Baroni}},
  \bibinfo {author} {\bibfnamefont {N.}~\bibnamefont {Bonini}}, \bibinfo
  {author} {\bibfnamefont {M.}~\bibnamefont {Calandra}}, \bibinfo {author}
  {\bibfnamefont {R.}~\bibnamefont {Car}}, \bibinfo {author} {\bibfnamefont
  {C.}~\bibnamefont {Cavazzoni}}, \bibinfo {author} {\bibfnamefont
  {D.}~\bibnamefont {Ceresoli}}, \bibinfo {author} {\bibfnamefont {G.~L.}\
  \bibnamefont {Chiarotti}}, \bibinfo {author} {\bibfnamefont {M.}~\bibnamefont
  {Cococcioni}}, \bibinfo {author} {\bibfnamefont {I.}~\bibnamefont {Dabo}},
  \bibinfo {author} {\bibfnamefont {A.}~\bibnamefont {{Dal Corso}}}, \bibinfo
  {author} {\bibfnamefont {S.}~\bibnamefont {de~Gironcoli}}, \bibinfo {author}
  {\bibfnamefont {S.}~\bibnamefont {Fabris}}, \bibinfo {author} {\bibfnamefont
  {G.}~\bibnamefont {Fratesi}}, \bibinfo {author} {\bibfnamefont
  {R.}~\bibnamefont {Gebauer}}, \bibinfo {author} {\bibfnamefont
  {U.}~\bibnamefont {Gerstmann}}, \bibinfo {author} {\bibfnamefont
  {C.}~\bibnamefont {Gougoussis}}, \bibinfo {author} {\bibfnamefont
  {A.}~\bibnamefont {Kokalj}}, \bibinfo {author} {\bibfnamefont
  {M.}~\bibnamefont {Lazzeri}}, \bibinfo {author} {\bibfnamefont
  {L.}~\bibnamefont {Martin-Samos}}, \bibinfo {author} {\bibfnamefont
  {N.}~\bibnamefont {Marzari}}, \bibinfo {author} {\bibfnamefont
  {F.}~\bibnamefont {Mauri}}, \bibinfo {author} {\bibfnamefont
  {R.}~\bibnamefont {Mazzarello}}, \bibinfo {author} {\bibfnamefont
  {S.}~\bibnamefont {Paolini}}, \bibinfo {author} {\bibfnamefont
  {A.}~\bibnamefont {Pasquarello}}, \bibinfo {author} {\bibfnamefont
  {L.}~\bibnamefont {Paulatto}}, \bibinfo {author} {\bibfnamefont
  {C.}~\bibnamefont {Sbraccia}}, \bibinfo {author} {\bibfnamefont
  {S.}~\bibnamefont {Scandolo}}, \bibinfo {author} {\bibfnamefont
  {G.}~\bibnamefont {Sclauzero}}, \bibinfo {author} {\bibfnamefont {A.~P.}\
  \bibnamefont {Seitsonen}}, \bibinfo {author} {\bibfnamefont {A.}~\bibnamefont
  {Smogunov}}, \bibinfo {author} {\bibfnamefont {P.}~\bibnamefont {Umari}}, \
  and\ \bibinfo {author} {\bibfnamefont {R.~M.}\ \bibnamefont {Wentzcovitch}},\
  }\href {\doibase 10.1088/0953-8984/21/39/395502} {\bibfield  {journal}
  {\bibinfo  {journal} {J. Phys. Cond. Matter}\ }\textbf {\bibinfo {volume}
  {21}},\ \bibinfo {pages} {395502} (\bibinfo {year} {2009})}\BibitemShut
  {NoStop}%
\bibitem [{\citenamefont {Kresse}\ and\ \citenamefont
  {Joubert}(1999)}]{Kresse1999}%
  \BibitemOpen
  \bibfield  {author} {\bibinfo {author} {\bibfnamefont {G.}~\bibnamefont
  {Kresse}}\ and\ \bibinfo {author} {\bibfnamefont {D.}~\bibnamefont
  {Joubert}},\ }\href {\doibase 10.1103/PhysRevB.59.1758} {\bibfield  {journal}
  {\bibinfo  {journal} {Phys. Rev. B}\ }\textbf {\bibinfo {volume} {59}},\
  \bibinfo {pages} {1758} (\bibinfo {year} {1999})}\BibitemShut {NoStop}%
\bibitem [{\citenamefont {Perdew}\ \emph {et~al.}(1997)\citenamefont {Perdew},
  \citenamefont {Burke},\ and\ \citenamefont {Ernzerhof}}]{Perdew1996}%
  \BibitemOpen
  \bibfield  {author} {\bibinfo {author} {\bibfnamefont {J.~P.}\ \bibnamefont
  {Perdew}}, \bibinfo {author} {\bibfnamefont {K.}~\bibnamefont {Burke}}, \
  and\ \bibinfo {author} {\bibfnamefont {M.}~\bibnamefont {Ernzerhof}},\ }\href
  {\doibase 10.1103/PhysRevLett.78.1396} {\bibfield  {journal} {\bibinfo
  {journal} {Phys. Rev. Lett.}\ }\textbf {\bibinfo {volume} {78}},\ \bibinfo
  {pages} {1396} (\bibinfo {year} {1997})}\BibitemShut {NoStop}%
\bibitem [{\citenamefont {Grimme}(2006)}]{Grimme2006}%
  \BibitemOpen
  \bibfield  {author} {\bibinfo {author} {\bibfnamefont {S.}~\bibnamefont
  {Grimme}},\ }\href {\doibase 10.1002/jcc.20495} {\bibfield  {journal}
  {\bibinfo  {journal} {J. Comp. Chem.}\ }\textbf {\bibinfo {volume} {27}},\
  \bibinfo {pages} {1787} (\bibinfo {year} {2006})}\BibitemShut {NoStop}%
\bibitem [{\citenamefont {Anisimov}\ \emph {et~al.}(1991)\citenamefont
  {Anisimov}, \citenamefont {Zaanen},\ and\ \citenamefont
  {Andersen}}]{Anisimov1991}%
  \BibitemOpen
  \bibfield  {author} {\bibinfo {author} {\bibfnamefont {V.~I.}\ \bibnamefont
  {Anisimov}}, \bibinfo {author} {\bibfnamefont {J.}~\bibnamefont {Zaanen}}, \
  and\ \bibinfo {author} {\bibfnamefont {O.~K.}\ \bibnamefont {Andersen}},\
  }\href {\doibase 10.1103/PhysRevB.44.943} {\bibfield  {journal} {\bibinfo
  {journal} {Phys. Rev. B}\ }\textbf {\bibinfo {volume} {44}},\ \bibinfo
  {pages} {943} (\bibinfo {year} {1991})}\BibitemShut {NoStop}%
\bibitem [{\citenamefont {Cococcioni}\ and\ \citenamefont
  {de~Gironcoli}(2005)}]{Cococcioni2005}%
  \BibitemOpen
  \bibfield  {author} {\bibinfo {author} {\bibfnamefont {M.}~\bibnamefont
  {Cococcioni}}\ and\ \bibinfo {author} {\bibfnamefont {S.}~\bibnamefont
  {de~Gironcoli}},\ }\href {\doibase 10.1103/PhysRevB.71.035105} {\bibfield
  {journal} {\bibinfo  {journal} {Phys. Rev. B}\ }\textbf {\bibinfo {volume}
  {71}},\ \bibinfo {pages} {035105} (\bibinfo {year} {2005})}\BibitemShut
  {NoStop}%
\bibitem [{\citenamefont {Dennis}\ and\ \citenamefont
  {Morέe}(1977)}]{Dennis1977}%
  \BibitemOpen
  \bibfield  {author} {\bibinfo {author} {\bibfnamefont {J.~E.}\ \bibnamefont
  {Dennis}}\ and\ \bibinfo {author} {\bibfnamefont {J.~J.}\ \bibnamefont
  {Morέe}},\ }\href@noop {} {\bibfield  {journal} {\bibinfo  {journal} {SIAM
  Rev.}\ }\textbf {\bibinfo {volume} {19}},\ \bibinfo {pages} {46} (\bibinfo
  {year} {1977})}\BibitemShut {NoStop}%
\bibitem [{\citenamefont {L{\"{o}}wdin}(1950)}]{Lowdin1950}%
  \BibitemOpen
  \bibfield  {author} {\bibinfo {author} {\bibfnamefont {P.-O.}\ \bibnamefont
  {L{\"{o}}wdin}},\ }\href {\doibase 10.1063/1.1747632} {\bibfield  {journal}
  {\bibinfo  {journal} {J. Chem. Phys.}\ }\textbf {\bibinfo {volume} {18}},\
  \bibinfo {pages} {365} (\bibinfo {year} {1950})}\BibitemShut {NoStop}%
\bibitem [{\citenamefont {Bader}(1990)}]{Bader1990}%
  \BibitemOpen
  \bibfield  {author} {\bibinfo {author} {\bibfnamefont {R.~F.~W.}\
  \bibnamefont {Bader}},\ }\href@noop {} {\emph {\bibinfo {title} {{Atoms in
  Molecules: A Quantum Theory}}}}\ (\bibinfo  {publisher} {Oxford University
  Press: Oxford, U.K.},\ \bibinfo {year} {1990})\BibitemShut {NoStop}%
\bibitem [{\citenamefont {Sugar}\ and\ \citenamefont
  {Musgrove}(1990)}]{Sugar1990}%
  \BibitemOpen
  \bibfield  {author} {\bibinfo {author} {\bibfnamefont {J.}~\bibnamefont
  {Sugar}}\ and\ \bibinfo {author} {\bibfnamefont {A.}~\bibnamefont
  {Musgrove}},\ }\href {\doibase 10.1063/1.555855} {\bibfield  {journal}
  {\bibinfo  {journal} {J. Phys. Chem. Ref. Data}\ }\textbf {\bibinfo {volume}
  {19}},\ \bibinfo {pages} {527} (\bibinfo {year} {1990})}\BibitemShut
  {NoStop}%
\bibitem [{\citenamefont {Wallace}(1947)}]{Wallace:PR1947}%
  \BibitemOpen
  \bibfield  {author} {\bibinfo {author} {\bibfnamefont {P.~R.}\ \bibnamefont
  {Wallace}},\ }\href {\doibase 10.1103/PhysRev.71.622} {\bibfield  {journal}
  {\bibinfo  {journal} {Phys. Rev.}\ }\textbf {\bibinfo {volume} {71}},\
  \bibinfo {pages} {622} (\bibinfo {year} {1947})}\BibitemShut {NoStop}%
\bibitem [{\citenamefont {McClure}\ and\ \citenamefont
  {Yafet}(1962)}]{MCCLURE196222}%
  \BibitemOpen
  \bibfield  {author} {\bibinfo {author} {\bibfnamefont {J.~W.}\ \bibnamefont
  {McClure}}\ and\ \bibinfo {author} {\bibfnamefont {Y.}~\bibnamefont
  {Yafet}},\ }in\ \href {\doibase
  http://dx.doi.org/10.1016/B978-0-08-009707-7.50008-X} {\emph {\bibinfo
  {booktitle} {Proceedings of the Fifth Conference on Carbon}}}\ (\bibinfo
  {publisher} {Pergamon},\ \bibinfo {year} {1962})\ pp.\ \bibinfo {pages}
  {22--28}\BibitemShut {NoStop}%
\bibitem [{\citenamefont {Hewson}(1997)}]{Hewson1993}%
  \BibitemOpen
  \bibfield  {author} {\bibinfo {author} {\bibfnamefont {A.~C.}\ \bibnamefont
  {Hewson}},\ }\href {\doibase 10.1017/CBO9780511470752} {\emph {\bibinfo
  {title} {The Kondo Problem to Heavy Fermions}}}\ (\bibinfo  {publisher}
  {Cambridge University Press},\ \bibinfo {year} {1997})\BibitemShut {NoStop}%
\bibitem [{\citenamefont {Ducastelle}(2013)}]{Ducastelle2013:PRB}%
  \BibitemOpen
  \bibfield  {author} {\bibinfo {author} {\bibfnamefont {F.}~\bibnamefont
  {Ducastelle}},\ }\href {\doibase 10.1103/PhysRevB.88.075413} {\bibfield
  {journal} {\bibinfo  {journal} {Phys. Rev. B}\ }\textbf {\bibinfo {volume}
  {88}},\ \bibinfo {pages} {075413} (\bibinfo {year} {2013})}\BibitemShut
  {NoStop}%
\bibitem [{\citenamefont {Kochan}\ \emph {et~al.}(2015)\citenamefont {Kochan},
  \citenamefont {Irmer}, \citenamefont {Gmitra},\ and\ \citenamefont
  {Fabian}}]{Kochan2015}%
  \BibitemOpen
  \bibfield  {author} {\bibinfo {author} {\bibfnamefont {D.}~\bibnamefont
  {Kochan}}, \bibinfo {author} {\bibfnamefont {S.}~\bibnamefont {Irmer}},
  \bibinfo {author} {\bibfnamefont {M.}~\bibnamefont {Gmitra}}, \ and\ \bibinfo
  {author} {\bibfnamefont {J.}~\bibnamefont {Fabian}},\ }\href {\doibase
  10.1103/PhysRevLett.115.196601} {\bibfield  {journal} {\bibinfo  {journal}
  {Phys. Rev. Lett.}\ }\textbf {\bibinfo {volume} {115}},\ \bibinfo {pages}
  {196601} (\bibinfo {year} {2015})}\BibitemShut {NoStop}%
\end{thebibliography}%

\end{document}